\documentclass[onecolumn,preprintnumbers,nofootinbib]{revtex4}
\usepackage{sidecap}
\usepackage{wrapfig}

\usepackage{graphicx}  
\usepackage{amssymb}
\usepackage{color}
\usepackage{enumerate}
\usepackage{amsmath}
\usepackage{slashed}
\usepackage{comment}
\def\beq{\begin{equation}}
\def\eeq{\end{equation}}
\def\beqn{\begin{eqnarray}}
\def\eeqn{\end{eqnarray}}

\def\cor{\color{red}}
\def\cob{\color{blue}}
\def\con{\color{black}} 

\renewcommand{\texttt}{{}}
\newcommand{\be}{\begin{eqnarray}}
\newcommand{\ee}{\end{eqnarray}}

\begin{document}

\title{
{
Finite Conformal Quantum Gravity 
and Nonsingular Spacetimes 
}}

\author{Leonardo Modesto}
\email{lmodesto@fudan.edu.cn}
\email{lmodesto1905@icloud.com}

\author{Les\l{}aw Rachwa\l{}}
\email{rachwal@fudan.edu.cn}

\affiliation{Center for Field Theory and Particle Physics and Department of Physics, \\
Fudan University, 200433 Shanghai, China   \\
}


\begin{abstract} \noindent

%
%
We explicitly prove that a class of finite quantum gravitational theories (in odd as well as in even dimension) is actually a range of anomaly-free conformally invariant theories in the spontaneously broken phase of the conformal Weyl symmetry. At classical level we show how the  
Weyl conformal invariance is likely able to tame the spacetime singularities that plague not only Einstein gravity, but also local and weakly non-local higher derivative theories. 
This latter statement is rigorously proved by a singularity theorem that applies to 
a large class of weakly non-local theories. 
Following the seminal paper by Narlikar and Kembhavi, we provide an explicit construction of 
singularity-free black hole exact solutions conformally equivalent to the Schwarzschild metric. 
Furthermore, we show that the FRW cosmological solutions and the Belinski, Khalatnikov, Lifshitz (BKL) spacetimes, which exactly solve the classical equations of motion, 
are 
conformally equivalent to regular spacetimes.
Finally, we prove that the Oppenheimer-Volkov gravitational collapse is a  
an exact (singularity-free) solution 
of the non-local conformally invariant theory compatible with the bounce paradigm.


%

\end{abstract}

\maketitle



\tableofcontents


\section{Introduction}

The problems of quantum gravity are long-standing and probably the most difficult problems of theoretical physics. Many ingenious ideas were proposed in order to find a fully consistent framework for it. One direction of these developments amounted to enlarging the symmetry group governing the gravitational dynamics. Obviously, this group should be fully gauged, so the new gauge symmetry should be realized in a local manner. The original hope was that the Ward identities of this new symmetry could constrain the quantum dynamics and hopefully provide more control over quantum divergences, which typically beset the quantum field theory of gravity. This was only partially successful with the inclusion of supersymmetry and the realization of supergravity models. However, there is still one symmetry, which was often neglected as not pertaining to our world. This is the conformal symmetry relating things at small and large scales. In the most naive version this symmetry was realized as invariance with respect to global transformations of the scale, according to the formula
\be
x^\mu\to\Omega \, x^\mu,
\ee
(here applied to controvariant dimensionful coordinates on flat Minkowski spacetime.) Later in a fully diffeomorphism (Diff.) covariant framework of general relativity (GR) this was promoted to a local transformation on the metric tensor according to the law
\be
g_{\mu\nu}(x)\to \Omega^2(x)g_{\mu\nu}(x). 
\ee
This transformation bears the name of Weyl rescaling and as it is clear it preserves only the angles (normalized scalar products of vectors), but not the spacetime distances (magnitudes of vectors), hence the other name of the symmetry is conformal. 

Since the scales are not absolute notions in conformally invariant theories, then it is possible to think that this symmetry may be instrumental in solving problems of quantum divergences and classical singularities in gravitational theories. Actually for the first part of the problem the conformal symmetry realized on the quantum level is the solution, because its presence (in both unbroken or spontaneously broken phase) is equivalent to the absence of all divergences. In this paper we make this argument precise and relate it on the other hand to the absence of conformal anomaly.

One of the problem with conformal quantum gravity was that it was very difficult to keep conformal invariance on all loop quantum levels, when it was secured on the classical level or to some lower loop levels. In other words this was seen as the problem with returning loop divergences appearing at every level due to non-renormalizability of quantum Einstein-Hilbert (E-H) gravity. Some control over divergences was gained in renormalizable gravitational theories with higher derivatives, however, the problem with unitarity spoiled the physical interpretation of them and rendered them inconsistent. The rescue to this situation came after the invention of weakly non-local gravitational 
theories \cite{Krasnikov, kuzmin,Tombo, Khoury, modesto,modestoLeslaw}. In this paper we briefly review them and expand about 
a range of them, in which they are unitary (ghost-free) and perturbatively super-renormalizable in the quantum field theory framework
\cite{Krasnikov, kuzmin,Tombo, Khoury, modesto,modestoLeslaw, universality, Briscese:2013lna, Cnl1, Dona,Mtheory,NLsugra,Modesto:2013jea}. 
Moreover, a recent mild extension of these theories has been proved to be completely 
finite at any order in the loop expansion \cite{modestoLeslaw}. In this way for the first time a class of quantum gravity theories was found completely free of any divergences. Therefore, they are candidates to be conformally invariant quantum gravity theories. Originally, they were not written in a form showing explicitly the conformal invariance and this is why later in the article, we propose a conformally invariant reinterpretation of these theories based on the following requirements:
(i) general covariance; 
(ii) explicit conformal invariance (in broken or unbroken phase); 
(iii) weak non-locality (or quasi-polynomiality); 
(iv) unitarity (freedom from ghosts) and
(v) finiteness at quantum level.  

In comparison with Einstein gravity we enlarge the symmetry group of gravitational dynamics by the inclusion of conformal invariance, which we believe to be crucial also in removing all kind of spacetime singularities in the classical physical solutions. 
At classical level in the wake of numerous approximate and exact 
``{\em singularity-free}" solutions \cite{ModestoMoffatNico,BambiMalaModesto2, BambiMalaModesto,calcagnimodesto, koshe1} we were led to believe that the non-locality in the kinetic terms for the fluctuations in the gravitational action 
was enough to solve the issue of spacetime singularities. 
However, in \cite{exactsol} we showed that all Einstein spaces, 
including the singular Schwarzschild and Kerr spacetimes \cite{exactsol}, 
are exact solutions of the non-local theory. 
Therefore, non-locality is not sufficient to remove the singularities and we need a new (actually old) symmetry principle to get rid out of them. 

Let us here bring up an analogy between conformal invariance and the role played by Diff. invariance in general relativity. It is well known that the Schwarzschild metric in Schwarzschild coordinates is singular at the event horizon. However, after years of
fighting experts of general relativity figured out that such singularity was not physical, but was just an artefact 
of an unlucky choice of coordinate system. Indeed, by changing coordinates to the Kruskal-Szekeres ones the singularity disappears and the spacetime can be easily extended beyond the event horizon. In this case singularity was of the coordinate type and was removed thanks to the invariance with respect to coordinate transformations.

We believe that in the same way, conformal invariance should remove all the spacetime essential singularities.
We may consider here a first simple example of spacetime singularities:
the initial Big Bang singularity in FRW models. The Weyl tensor for any FRW spacetime is exactly zero, which means that
in a conformally invariant theory the flat spacetime and any FRW spacetime are actually the same indistinguishable objects,
because they are in the same equivalence class of conformally related metrics. This is in the same way like different coordinate systems describe the same differential manifold, if they are related by a differentiable map in differential geometry. Phrasing differently FRW spacetime is conformally flat, because the tensor of the conformal curvature (Weyl tensor) vanishes there identically.
The FRW metrics are to the Weyl tensor like the Minkowski spacetime is to the Riemann tensor.
Therefore, there is no invariant physical content in the initial Big Bang singularity in these models in the same way as there is no physical content in the coordinate singularity at the event horizon of a Schwarzschild or Kerr back hole. 

These arguments are not very new, but actually there is an old \cite{Narlikar} and also 
new very inspiring literature about conformal gravity and its role in removing spacetime singularities \cite{thooft0,Sugget,thooft,barsTurok,Bars2}. 
However, our main contribution in this paper lies in proposing a conformally invariant theory that is finite at quantum level, and therefore, a theory that is devoid of any conformal anomaly. 
Moreover at classical level we have discovered a new class of exact singularity-free black hole solutions in a wide range of conformally invariant theories. These configurations are related by conformal transformations to the original singular Schwarzschild metric and constitute a core of the proposal for a conformal resolution of the black hole singularity.

Coming back to the hypotheses listed at the beginning of this section, the other difference with Einstein gravity lies in the third requirement from the above list, namely in the weak non-locality. It makes possible to achieve 
unitarity and finiteness at the same time of the full quantum theory. This is a solid statement confirmed by numerous studies \cite{kuzmin,Tombo,modesto, modestoLeslaw,universality}.

The paper is organized as follows. 

In the second section we remind and expand about a class of ghost-free weakly non-local gravitational theories. We explicitly compute the propagator for wide class of theories that mainly vary for the appearance or not of the Weyl tensor in the action. 
Power-counting super-renormalizability is in short proved. 
In the third section we present three range of weakly non-local theories in different basis:
Weyl, Bach, and Einstein. 
In section four we prove a simple but rigorous 
singularity theorem to be valid for a large class of weakly non-local theories ultraviolet complete. 
In section five we show the structure of the second variation of the action with respect to the graviton fluctuations in any dimension and we schematically display the one-loop counterterms. 
In the sixth section a range of classical conformally invariant actions is 
proposed, while 
in section seven such theories are studied at quantum level in odd dimension and for the particular case of four spacetime dimensions. It is shown, how the recently proposed finite quantum gravity can take the explicit form of a conformally invariant theory being, moreover, in its spontaneously broken phase. 
Finally, in sections eight, nine, and ten the spacetime singularities are taken by the horns and we get rid out of them with the help of the conformal symmetry on the footprint of the Narlikar and Kembhavi paper \cite{Narlikar}. As an example of singularity resolution we expressly construct a class of singularity-free spherically symmetric black hole solutions and BKL spacetimes. Finally, we study the gravitational collapse for dust matter. 
In particular in the sections nine and ten we explicitly show the geodesic completion of the above spacetimes making use of two different kind of probes: a massive particle and a conformally coupled 
particle. 
The FRW spacetime turns out to be automatically singularity-free in a conformally invariant theory because Weyl flat. 
At the end we write conclusions and speculate about possible applications of our finite conformal quantum gravity.

We wish to emphasize once again that all the results about the resolution of a wide class of spacetime singularities are general features of any conformally invariant theory. 
However, the quasi-polynomial theories are, at the moment, the only ones
compatible with quantum finiteness \cite{modestoLeslaw}, freedom from conformal anomaly, and perturbative unitarity \cite{Tombo}. For their application to cosmology of early universe we refer the reader to the recent paper \cite{KoshelevStaro}.


\section{The theory}\label{Theory}
The Lagrangian density of the most general $D$-dimensional theory weakly non-local (or quasi-local) and quadratic in the Riemann curvatures reads 
\cite{Krasnikov, Tombo, Khoury, modesto,modestoLeslaw, universality, Briscese:2013lna, Cnl1, Dona, Mtheory,NLsugra,Modesto:2013jea,M3,M4,entanglement}, 
\be
\mathcal{L}_{\rm g} = -  2 \kappa_{D}^{-2} \, \sqrt{g} 
\left( { R} 
+ {\bf Riem}  \, 
{\bf \gamma}(\Box) 
{\bf Riem} 
+ {V} 
\right)  
\equiv -  2 \kappa_{D}^{-2} \, \sqrt{g} \left(    R  + {R}_{\mu\nu\rho\sigma}  
\gamma(\Box)^{\mu \nu \rho \sigma}_{\alpha \beta \gamma \delta} \, 
{R}^{\alpha\beta\gamma\delta} 
+ { V} 
\right)   ,
\label{gravityG}  
\ee
where the weakly non-local function of the d'Alembertian operator $\gamma(\Box)$ is defined by
\be
\gamma(\Box)^{\mu \nu \rho \sigma}_{\alpha \beta \gamma \delta} =
 g^{\mu\rho} g^{\nu \sigma} g_{\alpha\gamma}  g_{\beta \delta} \gamma_0(\Box) +
  g^{\mu\rho} g_{\alpha\gamma} \delta^\nu_\beta \delta^\sigma_\delta \gamma_2(\Box)  
+ \delta^\mu_\alpha \delta^\nu_\beta \delta^\rho_\gamma \delta^\sigma_\delta \gamma_4(\Box).
\ee
%
%
The theory consists of a kinetic weakly non-local operator quadratic in the curvature, three entire functions 
$\gamma_0(\Box)$, $\gamma_2(\Box)$, $\gamma_4(\Box)$, and a potential ${ V}$, which we choose hereby to be local and at least cubic in the curvature.
In general dimension $D$ it is made up of the following three sets of operators, 
\be
 { V} = 
 \sum_{j=3}^{{\rm N}+2} \sum_{k=3}^{j} \sum_i c_{k,i}^{(j)} \left( \nabla^{2(j-k)} {\cal R}^k \right)_i
 \!
 \,+ 
 \sum_{j={\rm N}+3}^{\gamma+{\rm N}+1} \sum_{k=3}^{j} \sum_i d_{k,i}^{(j)} \left(\nabla^{2(j-k)} {\cal R}^k \right)_i 
 \,+ 
 \sum_{k=3}^{\gamma +{\rm N}+2} \!\!
 \sum_i s_{k,i} \, \left(  \nabla^{2 (\gamma + {\rm N}+2 -k )} \, {\cal R}^k \right)_i ,  
 \ee 
 where operators in the third set are called killers, because they are crucial in making the theory finite in any 
 dimension.
The coefficients,
 $c_{k,i}^{(j)}$, $d_{k,i}^{(j)}$, $s_{k,i}$ are coupling constants (only $c_{k,i}^{(j)}$ are undergoing RG running), while the tensorial structure of operators present in $V$ have been neglected\footnote{{\em Definitions ---} 
The metric tensor $g_{\mu \nu}$ has 
signature $(- + \dots +)$ and the curvature tensors are defined as follows: 
$R^{\mu}_{\nu \rho \sigma} = - \partial_{\sigma} \Gamma^{\mu}_{\nu \rho} + \dots $, 
$R_{\mu \nu} = R^{\rho}_{\mu  \rho \nu}$,  
$R = g^{\mu \nu} R_{\mu \nu}$. With symbol ${\cal R}$ we generally denote one of the above curvature tensors.}.
By 
$\Box = g^{\mu\nu} \nabla_{\mu} \nabla_{\nu}$ we denote the covariant box operator. 
The capital $\rm{N}$ is defined to be the following function of the spacetime dimension $D$:
$2 \mathrm{N} + 4 = D_{\rm odd} +1$ in odd dimensions and $2 \mathrm{N} + 4 = D_{\rm even}$
in even dimensions (in order to avoid fractional powers of the d'Alembertian operator).

The form factors $\gamma_i(\Box)$ must take the following particular forms, if we require the same spectrum as in the quantum Einstein-Hilbert gravity around Minkowski spacetime. We write them in terms of exponentials of entire functions $H_\ell(z)$ ($\ell=0,2$), namely 
\be
\hspace{-0.7cm} \gamma_0(\Box) = - \frac{(D-2) ( e^{H_0} -1 ) + D ( e^{H_2} -1 )}{4 (D-1) \Box} + \gamma_4(\Box) 
\label{gamma0} \, , 
\qquad 
\gamma_2(\Box) = \frac{e^{H_2} -1 }{\Box} - 4 \gamma_4(\Box) \, .
\label{gamma2}
\ee
The form factor $\gamma_4(\Box)$ stays arbitrary, but is only constrained by renormalizability to have the same (or lower in number of derivatives) asymptotic UV behaviour as the other two form factors $\gamma_\ell(\Box)$ ($\ell=0,2$). Due to dimensional reasons the form factor $\gamma_4(\Box)$ can be written as 
$\tilde{\gamma}_4(\Box)/\Box$, where now $\tilde{\gamma}_4(\Box)$ as well as $e^{H_0}$ and $e^{H_2}$ are dimensionless functions. 
The minimal choice compatible with unitarity and super-renormalizability corresponds to  
$\gamma_4(\Box) =0$.

As a matter of fact we can also add other operators quadratic in the curvature and equivalent to the above operators
up to interaction vertices. These operators correspond to a different ordering of derivatives in the form factors 
in-between the Riemann, Ricci, and scalar curvatures. We name these operators ``terminators".
However, such non-local operators can be crucial in making the theory finite \cite{kuzmin}, if 
we do not want to introduce any local (or non-local) term with more than two Riemann curvatures in the potential $\bf{V}$.
Here are some examples of terminators, 
\be
R \, \nabla_\alpha \frac{e^{H_{3}} -1}{\Box^2}\nabla^\alpha R,\quad
R_{\mu\nu}\nabla_\alpha\nabla_\beta \frac{e^{H_{4}} -1}{\Box^3} \nabla^{\alpha} \nabla^{\beta}R^{\mu\nu}, \quad 
  R_{\mu\nu\rho\alpha} \frac{e^{H_{5}} -1}{\Box^2} \nabla_{\beta}\nabla^{\alpha} 
 R^{\mu\nu\rho\beta} 
\, .
\ee

Finally, the entire functions $V^{-1}_{\ell}(z) \equiv \exp H_{\ell}(z)$ ($z \equiv - \Box_{\Lambda} \equiv - \Box/\Lambda^2$) ($\ell=0,2$) introduced in \eqref{gamma0}
are required to be real and positive on the real axis and without zeros on the 
whole complex plane $|z| < + \infty$. This requirement implies that there are no other
gauge-invariant poles than the transverse massless pole of the physical graviton (the same like in E-H theory).  We note that $\Lambda$ is an invariant mass scale in our fundamental theory, which later will be called the scale of non-locality. Moreover,  
%
there exists an angle $\Theta$ ($0<\Theta<\pi/2$ ), such that asymptotically
\be
|V^{-1}(z)| \rightarrow | z |^{\gamma + \mathrm{N}+1},\,\, {\rm when }\,\, |z|\rightarrow + \infty
\quad 
 {\rm with}%
\quad 
\gamma > \frac{D_{\rm even}}{2} 
\quad  {\rm or} 
\quad 
 \gamma > \frac{D_{\rm odd}-1}{2} \, , 
\label{tombocond}
\ee
for the complex values of $z$ in the conical regions $C$ defined by: 
$C = \{ z \, | \,\, - \Theta < {\rm arg} z < + \Theta \, ,  
\,\,  \pi - \Theta < {\rm arg} z < \pi + \Theta\}.$
The last condition is necessary to achieve the maximum convergence of the theory in
the UV regime and at the same time to avoid non-local counterterms.  One example of such function is:
\be
V^{-1}(z)= e^{\frac{1}{2} \left[ \Gamma \left(0, p(z)^2 \right)+\gamma_E  + \log \left( p(z)^2 \right) \right] } ,
\label{TomboFF}
\ee
where $p(z)$ is a polynomial of degree $\gamma+{\rm N} +1$. 
To achieve (super-)renormalizability 
the degrees of the polynomials appearing in the definitions of $V^{-1}_0(z)$ and $V^{-1}_2(z)$ must be equal. 
In the rest of the paper we will denote 
the common degree by $\gamma+{\rm N}+1$.

\subsection{Propagator and Unitarity}\label{propagatorS}
Now we want to obtain the propagator of the gravitational fluctuations around flat Minkowski background and discuss the issue of unitarity of the theory. Splitting the spacetime metric into the background and the fluctuation $h_{\mu \nu}$ 
defined by $g_{\mu \nu} =  \eta_{\mu \nu} + \kappa_D \, h_{\mu \nu}$,
we can expand the action (\ref{gravityG}) to the second order in $h_{\mu \nu}$.
The result of this expansion together with the usual harmonic gauge fixing term reads \cite{HigherDG}
$\mathcal{L}_{\rm quad} + \mathcal{L}_{\rm GF} = h^{\mu \nu} \mathcal{O}_{\mu \nu, \rho \sigma} \, h^{\rho \sigma}/2$,
where the operator 
$\mathcal{O}$ is made up of two terms, one coming from the quadratization of (\ref{gravityG})
and the other one from the following gauge-fixing term,
$\mathcal{L}_{\rm GF}  = \xi^{-1}  \partial^{\nu}h_{\mu \nu} \omega(-\Box_{\Lambda}) \partial_{\rho}h^{\rho \mu}$, where
$\omega( - \Box_{\Lambda})$ is a weight functional \cite{Stelle,shapiro3,Shapirobook}  and $\xi$ a gauge parameter.
The d'Alembertian operator in $\mathcal{L}_{\rm quad}$ and the gauge fixing term are written in terms of  
flat spacetime metric and partial derivatives.
Inverting the operator $\mathcal{O}$ \cite{HigherDG} and making use of the
form factors defined in (\ref{gamma0}), we find the 
two-point function in the harmonic gauge ($\partial^{\mu} h_{\mu \nu} = 0$),
\be
\mathcal{O}^{-1} = \frac{\xi (2P^{(1)} + \bar{P}^{(0)} ) }{2 k^2 \, \omega( k^2/\Lambda^2)} 
+ 
\frac{P^{(2)}}{k^2   e^{H_2(k^2/\Lambda^2)} }
\label{propagator} 
- \frac{P^{(0)}}{  \left( D-2 \right)k^2   e^{H_0(k^2/\Lambda^2)}} \,  .
\ee
%
Above we omitted the tensorial 
indices for the propagator $\mathcal{O}^{-1}$ and the usual projectors $\{ P^{(2)},P^{(1)},P^{(0)},\bar{P}^{(0)}\}$ 
are defined in  
\cite{HigherDG, VN}\label{proje2}.
%
We have also replaced $-\Box \rightarrow k^2$ in the quadratized action, thus writing it in momentum space.

The propagator (\ref{propagator}) describes the most general spectrum compatible with unitarity without any other degree of freedom besides the massless spin $2$ graviton field. We see that gauge-invariant are only terms proportional to $P^{(2)}$ and $P^{(0)}$. Unitarity is satisfied, because the propagator is given by multiplication of these two projectors by entire functions, respectively  $e^{H_2}$ and $e^{H_0}$, which do not give rise to any additional pole. Moreover, the optical theorem for the interaction between two gravitational sources $T_{1,2}(k)$ is trivially satisfied, namely 
 \be 
 2  \, {\rm Im} \left\{  T_1(k)^{\mu\nu} \mathcal{O}^{-1}_{\mu\nu, \rho \sigma} T_2(k)^{\rho \sigma} \right\} 
 =  2  \pi \con \, {\rm Res} \left\{  T_1(k)^{\mu\nu} \mathcal{O}^{-1}_{\mu\nu, \rho \sigma} T_2(k)^{\rho \sigma} \right\} \big|_{k^2 = 0}> 0 \, , 
 \label{optical}
 \ee
 where $T^{\mu\nu}(k)$ is the most general conserved energy-momentum tensor written in momentum space. 
 
 In the appendix  B we give two more examples of theories written in different bases, which nonetheless give rise to the same propagator as computed in this section.

Unitarity is proved by perturbing the Minkowski spacetime and the absence of ghosts and tachyons tells us about the stability of the flat spacetime. We can not exclude at the moment the presence of ghosts around 
other exact backgrounds. However, the theory is weakly non-local and the analysis developed in 
\cite{GhostsNL,GhostsNL2}
can be applied to infer that the lifetime of any background, on which ghosts could propagate, is not identically zero (unlike for example quadratic gravity \cite{Stelle}), but is always finite and can be arbitrary large depending on the 
particular exact solution  chosen \cite{Maggiore}. Moreover, once we give up locality for weak non-locality 
we do not have to worry about real ghosts in general backgrounds, because non-locality 
avoids the ``ghost catastrophe". 
This observation is quite remarkable, because it clarifies once and for all about the perturbative stability of gravitational fluctuations around any background. Indeed, the non-locality scale $\Lambda$ regularizes in a Lorentz-invariant way the decay probability \cite{GhostsNL,GhostsNL2,Maggiore} and the eventual presence of a real ghost tells about the lifetime of such spacetime, which can be very short or very long, but not identically zero, depending on the mass of the ghost developing around the peculiar background (the value of the mass can be read from the quadratic action and could be related to the mass scale $\Lambda$ and/or the form factor.) If there is such occurrence,  
unitarity is safe because the optical theorem can be satisfied whether it is taken the opposite prescription 
respect to normal particles to move out the real axis the pole, i.e
{\large
\begin{center}
\begin{tabular}{|c|c|c|}
\hline
\,
physical particle \,\,\, 
& \,   ghost (negative norm state)
& \, 
 \, ghost (positive  norm state) \\
\hline
\hline
\, $\frac{1}{i} \frac{1}{k^2 +m^2 - i \epsilon}$ \,   
 & $-\frac{1}{i} \frac{1}{k^2 + m^2 - i \epsilon}$ 
 & $- \frac{1}{i} \frac{1}{k^2 + m^2 + i \epsilon}$
\\
%
\hline
\end{tabular}
\label{good}
\end{center}
}
%
%
Indeed, we can calculate the scattering amplitude $T$, which is defined in terms of the $S-$matrix through the definition   
$S = 1+ i T$, to show that the optical theorem is satisfied, namely 
\be
\frac{-1}{k^2 + i \epsilon} =  {\rm P} \left( \frac{-1}{k^2} \right) + i \pi  \delta(k^2) \quad \Longrightarrow \quad 
2 {\rm Im} \, T_{if} = 2 {\rm Im} \left[ (-i) (-i)^2 \, \frac{1}{i} \frac{-1}{ k^2 + m^2 + i \epsilon} \right] 
\,\, \rightarrow \,\, 2 \pi \delta (k^2+m^2) \, .
\label{propdelta}
\ee
where the first factor $(-i)$ in the second equivalence above comes from the definition of $T$ and 
and the second $(-i)^2$ keeps track of the second order expansion of the $S-$matrix. The tensorial structure and the contraction of the propagator (\ref{propdelta}) with the energy-momentum tensor 
are the same of (\ref{optical}) 
 \cite{HigherDG}. 

The price to be paid is that 
the particles with negative energy are the ones that propagate forward in time, therefore the ghosts possess negative energy. 
This implies that in a scattering process involving normal particles and ghosts the energies of the normal particles can increase leading to a catastrophic instability of the vacuum (background) \cite{Cline}. 
However, in a non-local theory the catastrophe is avoided as explained above and in \cite{GhostsNL,GhostsNL2}. 

Let us consider 
a local theory consisting of a massless normal particle and a ghost-like particle with mass $m$.
The propagator reads,
\be
\frac{1}{k^2 \left( 1+ \frac{k^2}{m^2} \right) } = \frac{1}{k^2} - \frac{1}{k^2 + m^2}.
\ee
When we adopt the unitary prescription to avoid the pole in $k^2=-m^2$ the 
high energy convergence of the theory due to the propagator scaling $\sim1/k^4$ is spoiled.
This is due to the difference between the convergent propagator and the unitary one \cite{Stelle}, namely 
\be
\frac{1}{k^2 - i \epsilon} - \frac{1}{k^2 + m^2 - i \epsilon } - \left(  
\frac{1}{k^2 - i \epsilon} - \frac{1}{k^2 + m^2 + i \epsilon }
\right) = 
- 2 i \pi \, \delta(k^2+m^2) \, . 
\ee
When the extra $\delta(k^2+m^2)$ is integrated inside loop diagrams new non-renarmalizable divergences 
are generated \cite{Stelle}. 
The same argument applies to a weakly non-local theory where likely
the distribution $\delta(k^2+m^2)$ is replaced by $e^{- H(k^2)} \delta(k^2+m^2)$, which is actually equivalent to the distribution $e^{- H(m^2)} \delta(k^2+m^2)$. 

We end up with a non-renormalizable theory when we quantize the action 
around an unstable shortly- or longly-lived vacuum. The lifetime of the vacuum sets 
the shorter distance that can be measured and the effective action consists of 
a finite number of relevant operators respect up to such physical cut-off scale. Indeed, the unstable vacuum decays directly, or indirectly to a stable one (for example the Minkowski vacuum) 
before the effects of an infinite tower of operators characterizing the non-renormalizable action could be observed. In other words the theory has a physical cut-off.

Finally, if the soft non-locality at high energy invoked in \cite{Cline} is not enough to make finite the lifetime 
of the background manifold, then there is no problem to deal anyway, because unitarity is preserved, and  
such spacetime simply does not physically exist. 
Nevertheless, it is an exact solution of the classical EOM and only exists as a mere mathematically solution.

\subsection{Super-renormalizability and Finiteness}
%
 %
We now review the power-counting analysis of quantum divergences. 
In the high energy regime, 
the above propagator (\ref{propagator}) in momentum space 
scales schematically  as: 
$\mathcal{O}^{-1}(k) \sim k^{- (2 \gamma +D) }$.  
The vertices can be collected in different sets that may or may not involve the entire functions $\exp H_\ell(z)$. 
However, to find a bound on the quantum divergences it is sufficient to concentrate on
the leading operators in the UV regime. 
These operators scale as the propagator giving the following 
upper bounds on the superficial degree of divergence of any graph $\omega(G)$ in even dimension \cite{modesto,A}, 
\be
&& \delta^D(K) \, \Lambda^{2 \gamma(L-1)} \int (d^D p)^L \left( \frac{1}{p^{2 \gamma+D}} \right)^I \left( p^{2 \gamma +D} \right)^V 
 =
\delta^D(K) \, \Lambda^{2 \gamma(L-1)} \int (d^D p)^L \left( \frac{1}{p^{2 \gamma+D}} \right)^{L-1} 
\nonumber \\
&&
= \delta^D(K) \, \Lambda^{2 \gamma(L-1)} \, \left( \Lambda_{\rm cut-off}\right)^{\omega(G)} \,  ,  \qquad 
\boxed{\omega(G) \equiv D - 2 \gamma  (L - 1)} \,\, ,
\label{PC}
\ee
%
where we have introduced the following notation: $V$ for the 
numbers of vertices, $I$ for the number of internal lines, $L$ for the 
number of loops, $K$ for the sum of external momenta, $\Lambda_{\rm cut-off}$ for the cut-off scale. We also used the topological relation valid for any graph: $I = V + L -1$. 
Thus, if $\gamma > D/2$, only 1-loop divergences survive.  
Therefore, in even dimension
the theory is super-renormalizable \cite{kuzmin, Tombo, modesto, Krasnikov, A}
and only a finite number of operators of mass dimension up to $D$ has to be
included in the action in the renormalization procedure. 
In spacetimes of odd dimension we have defined $D_{\rm odd} +1 = 2 {\rm N} + 4$ and therefore 
\be
\omega(G)_{\rm odd} = D_{\rm odd} - (2 \gamma+1)  (L - 1), 
\ee
and for $\gamma > (D_{\rm odd}  - 1)/2$ there are at most one loop divergences. However, 
in odd dimension we can not construct any curvature invariant with an odd number of derivatives and the theory is completely finite (see also (\ref{gammaDiv})). 

Finally, a restricted number of operators in the potential ${ V}$ is enough to get rid out of the one loop divergences in even dimension.
In $D=4$ and using the dimensional regularization scheme (DIMREG) two operators quartic in the Riemann or Weyl tensors are 
sufficient to end up with all beta functions identically zero, namely 
\be
{V}({ R}) = s^{(1)}_{\rm R}  R_{\mu\nu} R^{\mu\nu} \Box^{\gamma -2} 
R_{\alpha \beta} R^{\alpha \beta} +
s^{(2)}_{\rm R}  R^2  \Box^{\gamma -2} R^2 \, , 
\label{KR}
\ee
or in terms of Weyl tensor solely 
\be
{V}({\bf C}) = s^{(1)}_{\rm C}  C_{\mu\nu\rho\sigma} C^{\mu\nu\rho\sigma} \Box ^{\gamma -2} C_{\alpha \beta \gamma \delta}
C^{\alpha \beta \gamma \delta} +
s^{(2)}_{\rm C}  C_{\mu\nu\rho\sigma} C^{\alpha \beta \gamma \delta} \Box ^{\gamma -2} C_{\alpha \beta \gamma \delta}
C^{\mu\nu\rho\sigma} \, . 
\label{KW}
\ee
These operators give contributions to the beta functions for ${ R}^2$ and ${\bf Ric}^2$  
linear in the front coefficients $s^{(1)}_{\rm R}$, $s^{(2)}_{\rm R}$ or 
$s^{(1)}_{\rm C}$, $s^{(2)}_{\rm C}$. Therefore, we can aways make zero the beta function with a suitable choice of the non-running coefficients. 
This is evident in the background field method because at one loop we need the second order variation of the action respect to $h_{\mu\nu}$, and such variation 
can 
is at least quadratic in the curvature for the operators  
(\ref{KR}) and (\ref{KW}) because they are quartic in the Riemann tensor. 
Contributions to the second order variation higher in curvature do not affect the beta functions 
$\beta_{ R^2}$ and 
$\beta_{\bf Ric^2}$. 
See the references \cite{modesto,modestoLeslaw, universality} for more details 
about super-renormalizability, finiteness, and killer operators.
%



\section{
Theories in different bases}
In this section we consider three 
weakly non-local gravitational actions out of the general ones (\ref{gravityG}) introduced in section \ref{Theory}. All these theories will contain three operators: the Einstein-Hilbert local operator and two
non-local operators quadratic in the curvature. 
The first theory is quadratic in the Ricci tensor and the Weyl tensor,
the second one is quadratic in the Ricci scalar and the Bach tensor, the third one is quadratic 
in the Ricci tensor and Ricci scalar.  All these theories are ghost free, perturbative unitary, and 
super-renormalizable or finite at quantum level. 

\subsection{The theory in Weyl basis}
We hereby define a class of theories in the Weyl basis. 
These theories are equivalent to the previous ones (\ref{gravityG}) for 
everything about unitarity (the propagator is given again by (\ref{propagator})) and super-renormalizability. Also finiteness can be easily gotten, if we slightly modify the killer operator terms.
The general Lagrangian density reads,
\be
&& \mathcal{L}_{\rm C} = -  2 \kappa_D^{-2} \sqrt{|g|}  
\Big[ { R} +
 {\bf C} 
  \gamma_{\rm C} (\Box) {\bf C}
 + { R}  \gamma_{\rm S}(\Box) { R } 
 + {\bf Riem} \,  \gamma_{\rm R}(\Box) {\bf Riem} + { V}({\bf C})
 \Big] , 
 \label{TWeyl}
 \ee
 with form factors defined by
 \be
\gamma_{\rm C} = - \frac{D-2}{4}  \gamma_2 \, , \quad 
\gamma_{\rm S} = \gamma_0  +  \frac{1}{2(D-1)}  \gamma_2 
 \quad \rm{and} \quad
 \gamma_{\rm R} = \gamma_4 + \frac{D-2}{4}  \gamma_2 \, ,
\ee
where all the form factors $\gamma_{\ell}$ $(\ell=0,2)$ are defined in (\ref{gamma2}). Solving for
$\gamma_{\rm C}$, $\gamma_{\rm S}$ and $\gamma_{\rm R}$ we find 
\be
&& 
\gamma_{\rm C} = \frac{(2 -D) \left(e^{H_2} -1 - 4 \gamma _4 \Box \right)}{4 \Box}
  ,  \\
&&  
\gamma_{\rm S} =\frac{(2-D) \left(e^{H_0}+e^{H_2}-2\right)+4 \gamma _4 (D-3) \Box }{4 (D-1) \Box} ,
\nonumber \\
&& 
\gamma_{\rm R} = 
\frac{(D-2) \left(e^{H_2}-1\right)-4 \gamma _4 (D-3) \Box}{4 \Box}  .
\ee
For the sake of simplicity we can assume 
$\gamma_{\rm R} =0$, and the theory (\ref{TWeyl}) reduces to
\be
\hspace{-0.2cm}
{\mathcal{L}_{\rm C} \label{TWeyl2} 
=- 2 \kappa_D^{-2}  \sqrt{|g|}\Big[ {R} + 
 {\bf C} 
  \gamma_{\rm C} (\Box) {\bf C}
  + { R}  \gamma_{\rm S}(\Box) { R }
  + { V}({\bf C})
 \Big] } \,  , \,\,
\gamma_{\rm C} =   \frac{D-2}{4(D-3)}  \frac{e^{H_2} -1}{\Box} \, , \,\,
\gamma_{\rm S} = - \frac{D-2}{4(D-1)}  \frac{e^{H_0} -1}{\Box}  .
\ee
Further specifications of this theory are possible in order to achieve finiteness at quantum level.  
It is enough to include a curvature potential $ V$ that is built up with only Weyl tensors, namely ${ V}({\bf C})$. This can always be done as explained at the end of section in section (\ref{Theory}) 
(see formulas (\ref{KW}).)
%

 For the theory written in the Weyl basis as presented here (\ref{TWeyl2}), the FRW metric for conformal matter ($T_{\rm matter} \equiv 0$, where $_{\rm matter}$ is the trace of the matter energy tensor) 
 solves exactly Einstein EOM 
 as well as the non-local EOM. In conclusion the Big-Bang singularity shows up in exact solutions of our 
 finite theory of quantum gravity. 
 However, if the gravitational sector also enjoys conformal invariance, then the FRW singular spacetime is conformally equivalent to the flat spacetime (because we can perform a conformal rescaling) and therefore the singularity is unphysical as extensively explain in the second part of this paper. 

 \subsection{The four dimensional theory in Bach basis}
 In this section we publish for the first time a super-renormalizable or finite theory of gravity in $D=4$
 making use of the Ricci scalar and the Bach tensor that is defined by
 \be
&& B_{a c} = \nabla^b \nabla^d C_{abcd} - R^{bd} C_{abcd} \label{Bac} \\
 && \hspace{0.67cm}
 =  \frac{1}{2} \Box R_{ac} - \frac{1}{12} g_{ac} \Box R - \frac{1}{6} \nabla_c \nabla_a R
 - R^{bd} C_{a b c d} - \frac{1}{3} R R_{a c} - \frac{1}{4} g_{a c} R_{b d} R^{b d} + \frac{1}{12} g_{ac} R^2 
\, .
 \ee
 The Bach tensor has zero divergence and conformal weight $-2$, i.e.
 \be
 \nabla^c B_{ac  } =0 \, , \quad B^{\Omega}_{ab} = \Omega^{-2} B_{ab} \, . 
 \ee
 Notice that $B_{ab}$ is identically zero for Ricci flat and FRW metrics. 
  The four dimensional Lagrangian reads
 \be
\hspace{-0.2cm}
{\mathcal{L}_{\rm B} \label{TWeyl3} 
=- 2 \kappa_4^{-2}  \sqrt{|g|}\Big[ { R} + 
 {\bf B} 
  \gamma_{\rm B} (\Box) {\bf B }
  + { R}  \gamma_{\rm S}(\Box) { R} 
  + { V}({\bf B})
 \Big] } \,  , \,\,
\gamma_{\rm B} =   4  \frac{e^{H_2} -1}{\Box^3} \, , \,\,
\gamma_{\rm S} = - \frac{1}{6}  \frac{e^{H_0} -1}{\Box}  .
\ee

It is straightforward to see from the first variation of the action that all Ricci flat spacetimes and the FRW spacetimes, whether they are sourced by a traceless 
energy tensor, are exact solutions of the classical equations of motion. 
This completes and makes even stronger the claim in the paper \cite{exactsol}, namely we have here an example of finite quantum gravity with the same black hole and cosmological Big Bang singularities of Einstein gravity present in exact solutions on classical level.  
  Therefore, the non-local smearing of the source, so successful in removing the Newtonian singularity,
has no general validity, and
only a new symmetry can definitely remove the spacetime singularities. 

The above claim will be rigorously proved later in this paper making use of a simple theorem. 


We can use the tensor defined in (\ref{Bac}) also in any dimension, but the conformal properties 
of $B_{ab}$ are not preserved in extra dimension.  
The multidimensional theory reads\footnote{A useful formula we made use to compute the propagator is:
\be
&& B_{a c} = \nabla^b \nabla^d C_{abcd} - R^{bd} C_{abcd}  = \frac{D-3}{D-2} \Box R_{a c} 
- \frac{D-3}{2 (D-1)} R_{;a c} + \frac{D (D-3) }{(D-2)^2} R_{ab} R^b_c 
- \frac{D-3}{D-2} R^{bd} C_{a b c d} - \frac{D-3}{ (D-2)^2} g_{a c} R_{b d} R^{bd }
\nonumber \\
&& \hspace{4.1cm}
- \frac{D (D-3) }{ ( D-1) (D-2)^2} R R_{ ac }
- \frac{D-3}{ 2 (D-1) (D-2)} g_{a c} \Box R
+ \frac{D-3}{ (D-1) (D-2)^2} g_{a c} R^2 
- R^{bd} C_{abcd}\, .
\label{BACHD}
\ee },
\be
&& 
{\mathcal{L}_{\rm B} \label{TWeyl3D} 
=- 2 \kappa_D^{-2}  \sqrt{|g|}\Big[ { R} + 
 {\bf B} 
  \gamma_{\rm B} (\Box) {\bf B }
  + { R}  \gamma_{\rm S}(\Box) { R} 
  + { V}({\bf B})
 \Big] } \,  , \,\,
 \\ 
&& 
\gamma_{\rm B} =   \left( \frac{D-2}{D-3}\right)^2  \frac{e^{H_2} -1}{\Box^3} \, , \quad 
\gamma_{\rm S} = 
-  \frac{(D-2)(D-3) \left( e^{H_0} -1 \right)  + D (D-4) \left( e^{H_2} -1 \right)}{ 4(D-1) (D-3) \Box}  \, .
\ee
Notice that all the Ricci flat and the FRW spacetimes for the case of traceless matter are exact solutions in extra dimension too.

 \subsection{The theory in Einstein basis}
Last but not least, we express the theory in the original basis introduced in \cite{Tombo, modesto}.
For the sake of simplicity we assume $\gamma_4 =0$ in (\ref{gravityG}),
\be
\mathcal{L}_{\rm E} 
=  - \frac{ 2}{ \kappa_D^{2} } 
\sqrt{|g|}\Big[ {R} +
 {\bf G} 
\,   \gamma_{\rm G} (\Box) {\bf Ric}
 + { R}  \gamma_{\rm S^{\prime}}(\Box) { R }
  + {V} 
 \Big] ,
 \ee
 with form factors given by
 \be
 \gamma_{\rm G} =  \gamma_2 =  \frac{e^{H_2} -1 }{\Box} \, , \quad  \label{gammaG} 
\gamma_{\rm S^{\prime}} = \frac{1}{2} \gamma_2 + \gamma_0 = \frac{D-2}{4(D-1)} \frac{e^{H_2} - e^{H_0}}{\Box} \, . 
\label{EinsteinBasis}
\ee
We can make the further minimal choice $H_0=H_2$ and the theory reduces to
\be
&& \hspace{-0.5cm} \mathcal{L}_{\rm E} = -  2 \kappa_D^{-2} \sqrt{|g|}\Big[ { R} +
 {\bf G} 
\,   \gamma_{\rm G} (\Box) {\bf Ric}
  + { V}
 \Big]  \,  ,
\label{EinsteinBasisS}
\ee
where $\gamma_{\rm G}$ is given in (\ref{gammaG}). Notice that for $H_0=H_2$ the spacetime dimension 
$D$ disappears from the action.

\section{Singularity Theorem in non-local gravity}
Throughout all the paper we pointed out more times that almost on all spacetime singularity 
of Einstein gravity remain in non-local theories. In this section we explicitly prove 
a simple singularity theorem based on the very general theory (\ref{TWeyl3D}).


{\bf Theorem}. Given the EOM $E_{\mu\nu} = 8 \pi G_N T_{\mu\nu}$ 
for the theory (\ref{TWeyl3D}), which we derive below in a very compact form, 
the following implication turns out to be true,
\begin{itemize}
\item
$R_{\mu\nu} = 0 \quad  \&  \quad T_{\mu\nu} =0 \quad \Longrightarrow \quad E_{\mu\nu} = 0$ ; 
\item 
$R_{\mu\nu} = 8 \pi G_N T_{\mu\nu}\quad   \& \quad T^{\mu}_{\mu} = 0 \quad \Longrightarrow \quad E_{\mu\nu} = 8 \pi G_N T_{\mu\nu}$ .
\end{itemize}
Therefore, 
\begin{itemize}
\item all Ricci flat spacetimes are exact solutions of the theory (\ref{TWeyl3D}); 
\item all the FRW spacetime sourced by conformally coupled matter are exact solutions of the theory 
(\ref{TWeyl3D}). 
\end{itemize}
The theorem works for any potential at least quadratic in the Bach tensor. 

{\bf Proof.}
Let us start writing the exact EOM in a short and very compact notation, namely 
\be
&& \hspace{-0.5cm}
E_{\mu\nu} 
 = \frac{ \delta \left[  \sqrt{|g|} \left( R 
 + R \gamma_{\rm S}(\Box) R+B_{\alpha \beta} 
\gamma_{\rm B} (\Box) B^{\alpha \beta} + { V( B ) } \right) \right]}{\sqrt{|g|} \delta g^{\mu\nu}} \nonumber \\
 && \hspace{-0.5cm}
 = G_{\mu\nu}  -  
 \frac{1}{2} g_{\mu\nu} \left(R \gamma_{\rm S} (\Box) R \right) -  
 \frac{1}{2} g_{\mu\nu} \left(B_{\alpha \beta} \gamma_{\rm B} (\Box)
 B^{\alpha \beta} \right) 
 +   2\frac{\delta R}{\delta g^{\mu \nu}  } \left( \gamma_{\rm S} (\Box)
R \right)  
+ \frac{\delta B_{\alpha \beta}}{\delta g^{\mu \nu}  } 
\left( \gamma_{\rm B} (\Box) B^{\alpha \beta} \right) 
+   \frac{\delta B^{\alpha \beta}}{\delta g^{\mu \nu}  } \left( \gamma_{\rm B}(\Box)
 B_{\alpha \beta} \right)  \nonumber \\
  && \hspace{-0.1cm}
 +  \frac{\delta \Box^r}{\delta g^{\mu\nu} }
 \left( 
\frac{  \gamma_{\rm S}(\Box^l)
 -\gamma_{\rm S}(\Box^r) 
}{\Box^l - \Box^r} 
 R R \right) 
+  \frac{\delta \Box^r}{\delta g^{\mu\nu} }
 \left( 
  \frac{ \gamma_{\rm B}(\Box^l)
- \gamma_{\rm B}(\Box^r)
}{\Box^l - \Box^r} 
 B_{\alpha \beta} B^{\alpha \beta} \right) 
 + 
 \frac{1}{\sqrt{|g|} } 
 \frac{ \delta    { V( B ) }   }{ \delta g^{\mu\nu} } 
 = 8 \pi G_N T_{\mu\nu}  \, ,
 \label{EOMnl}
\ee
where $\Box^{l,r}$ act on the left and right arguments (on the right of the incremental ratio) as indicated inside the brackets. 

We observe that when we replace $R_{\mu\nu} =0$ and $T_{\mu\nu} = 0$  
in the above EOM (\ref{EOMnl}) the tensor $E_{\mu\nu}$ is identically zero.
Indeed, the EOM contain operators linear and quadratic respectively in the Ricci or the Bach tensor 
and the Bach tensor vanishes when ${\bf Ric} = 0$. Therefore, the first item of the theorem is proved. 
We now proceed with the second item. 

For the FRW spacetimes the Bach tensor is identically zero (see the definitions (\ref{Bac}) and (\ref{BACHD})) and when radiation 
(or general conformal matter) is coupled to the gravitational theory (\ref{TWeyl3D}) only the term quadratic in the Ricci scalar and the Einstein-Hilbert term in the action (\ref{TWeyl3D}) 
give contribution to the EOM. 
This is clear looking at the EOM ({\ref{EOMnl}). Indeed, all the terms containing the Bach tensor 
(without variation respect to the metric) are identically zero for any FRW spacetime. 
Therefore, the EOM simplify to 
\be
   G_{\mu\nu} 
- \frac{1}{2} g_{\mu\nu} \left(R \gamma_{\rm S}(\Box) R \right) 
+   2   \frac{\delta R}{\delta g^{\mu \nu}  } \left( \gamma_{\rm S} (\Box)
R \right)
+  \frac{\delta \Box^r}{\delta g^{\mu\nu} }
 \left( 
\frac{  \gamma_{\rm S}(\Box^l)
 -\gamma_{\rm S}(\Box^r) 
}{\Box^l - \Box^r} 
 R R \right) 
= T_{\mu\nu} \, . 
 \label{EOMnlC2}
\ee
We now evaluate the trace of the above EOM (\ref{TraceEOMnlC}),
\be
\hspace{-0.4cm}
   R  + \frac{D}{2}  R \gamma_{\rm S} (\Box) R 
-   2 g^{\mu\nu} \frac{\delta R}{\delta g^{\mu \nu}  } \left( \gamma_{\rm S}(\Box)
R \right)
 -  g^{\mu\nu}
 \frac{\delta \Box^r}{\delta g^{\mu\nu} } \!
 \left( 
\frac{  \gamma_{\rm S}(\Box^l)
 -\gamma_{\rm S}(\Box^r) 
}{\Box^l - \Box^r} 
 R R \right) 
\! = 0  .
 \label{TraceEOMnlC}
 \ee
Notice that the right hand side is identically zero because for radiation 
(and any other conformal matter coupled to gravity) the trace of the energy tensor vanishes. 
The trace of the EOM is solved by $R=0$ that we can replace in (\ref{EOMnlC2}) and we finally end up with the Einstein equations. Therefore, we end up with exactly the same reduced EOM as for 
Einstein gravity and the usual Big Bang singular solution for a radiation dominated universe is unavoidable. 

In short we can summarize the proof of the second item with the following chain of implications, 
\be
\hspace{-0.5cm}
 {\rm tr\, {\bf{T}} } = 0 \ \Longrightarrow \ 
  {R} = 0 \  \Longrightarrow \ {\bf G} = 8 \pi G_N\, {\bf T}  \,\,
\Longrightarrow \,\, \boxed{\mbox{Big Bang Singularity} }\, .
\label{FRWbb}
\ee 
At this level we have proved that the spacetime singularities show up also in a finite theory of quantum gravity. We believe that only a fundamental symmetry principle may sweep away the spacetime 
singularities. 

\section{Local higher derivative quantum gravity} \label{prototype}
In this technical section we study a quite general local super-renormalizable quantum gravity.
The results will be exported later to our unitary weakly non-local super-renormalizable gravitational theory.
For our goals the theory in this section can be consider as a  kind of prototype. 
This is possible, because the local and weakly non-local theories under investigation have the same divergences and therefore the same beta functions. 

Let us start with the following general prototype for a local super-renormalizable action, 
\be
&& 
\hspace{-1.2cm}
 S_{\rm HD} 
  = \int d^D x \sqrt{|g|} \left[  
\bar{\lambda} - 2 \kappa_D^{-2}  R 
+ 2 \tilde{\kappa}_D^{-2} \,  \sum_{n=0}^{\gamma+{\rm N}}  \omega_{{\rm Ric}, n} \, 
  R_{\mu \nu} \, \Box^{n}  R^{\mu \nu}  
+ 2 \tilde{\kappa}_D^{-2} 
\,  \sum_{n=0}^{\gamma+ {\rm N}}  \, 
 \omega_{{\rm R}, n} \,
R \,  \Box^n R \right] . 
     \label{ActionUV2}
\ee
In background field method the metric $g_{\mu\nu}$ is split into a background metric $\bar{g}_{\mu\nu}$ 
and a quantum fluctuation $h_{\mu\nu}$, 
\be
g_{\mu\nu} = \bar{g}_{\mu\nu} + h_{\mu \nu}. 
\label{BFM}
\ee
Sometimes below we will denote these metrics by $g$, $\bar{g}$ and $h$ without writing covariant indices explicitly. Additionally from now on we will not speak, in this section, about the full metric $g$ and for simplicity of notation the background metric will be denoted again by $g$. 
Since the theory is diff. invariant we have to fix the gauge and in the quantization procedure we must introduce  Faddeev-Popov (FP) ghosts.
The gauge-fixing and FP-ghost actions read as follows, 
\be
&& \hspace{-1.2cm} 
S_{\rm gf}  = \int \! d^D x \sqrt{ |g| } \, \frac{1}{2}\, \chi_{\mu} \,  C^{\mu \nu} \, \chi_{\nu} \, , 
\quad \chi_{\mu} = {\nabla}_{\sigma} h^{\sigma}_{\mu} - \beta_g  {\nabla}_{\mu} h \, , 
\quad 
C^{\mu \nu} \! = - \frac{1}{\alpha_g}
 \left( {g}^{\mu \nu} \Box 
 + \gamma_g {\nabla}^{\mu} {\nabla}^{\nu} - {\nabla}^{\nu} {\nabla}^{\mu} \right)  
 \Box_{\Lambda}^{  {\rm N} + \gamma}  , \label{shapiro3ain} \\
&& \hspace{-1.2cm}  
S_{\rm gh} = \int \! d^D x \sqrt{ |{g}| } \left[  \bar{C}_{\alpha} \, M^{\alpha}{}_{\beta} \, C^{\beta} 
+\frac{1}{2}  b_{\alpha} C^{\alpha \beta} b_{\beta} \right] \, , 
\quad 
M^{\alpha}{}_{\beta} = \Box \delta^{\alpha}_{\beta}  
+{\nabla}_{\beta} {\nabla}^{\alpha}  - 2 \beta_g {\nabla}^{\alpha} {\nabla}_{\beta} . 
\label{shapiro3a}
\ee
In (\ref{shapiro3ain}) and \eqref{shapiro3a}
we used covariant  
gauge-fixing condition $\chi_\mu$ with weight function $C^{\mu\nu}$  \cite{shapiro3} (the special case of it is an $\omega(\Box)$ appearing in the harmonic gauge fixing as introduced in section (\ref{propagatorS})). The standard (complex) FP-ghost and anti-ghost fields we denote by $C^{\beta} $ and $\bar{C}_{\alpha}$ respectively. Due to the higher derivative character of our theory we are forced to introduce also a third (real-)ghost field \cite{Shapirobook}, which we appoint $b_\alpha$. 
The gauge-fixing parameters
$\beta_g$ and $\gamma_g$ are dimensionless,  while $[\alpha_g] = M^{4-D}$. 
We notice right here that in our theory the beta functions are independent of these gauge parameters   
(see \cite{shapiro3} for a rigorous proof).

The partition function  of the full quantum theory with the right functional measure 
compatible with BRST invariance \cite{Anselmi:1991wb, Anselmi:1992hv, Anselmi:1993cu} reads 
\be
&& \hspace{-0.5cm}
Z[g] = \int \!  
\mu(g, h )
\prod_{\mu \leqslant \nu} \mathcal{D}h_{\mu\nu} \prod_{\alpha} \mathcal{D} \bar{C}_{\alpha}
 \prod_{\beta}  \mathcal{D}{C}^{\beta}
 \prod_{\gamma}  \mathcal{D}{b}_{\gamma} \, 
 e^{i \left[  S_{\rm HD} +  S_{\rm gf} + S_{\rm gh}   \right]} \,  .
\label{QG}
 \ee
 At one loop we can evaluate the functional integral explicitly and express the partition function as a product of functional determinants,
 namely
  \be
  Z[g] = e^{i S_{\rm g}[{g}]}   \left\{   \left. {\rm Det}\left[ 
   \frac{\delta^2 (S_{\rm HD}[g+h] + S_{\rm gf}[g+h]) }{\delta h_{\mu \nu} \delta h_{\rho \sigma}} \right\vert_{h=0} \right] \right\}^{-\frac{1}{2}} 
 ({\rm Det} \, M^{\alpha}{}_{\beta} ) \, \, 
({\rm Det} \, C^{\mu \nu} )^{\frac{1}{2}}  \, . 
 \nonumber 
 \ee
 By symbol $S_{\rm g}[g]$ we understand classical functional of the gravitational action of the theory for the background metric $g$.
To calculate the one-loop effective action we need first to expand the action plus the gauge-fixing term to the second order in the quantum fluctuation $h_{\mu \nu}$
\be
 \hat{H}^{\mu \nu , \rho \sigma}= \frac{\delta^2 S_{\rm HD}}{ \delta h_{\mu \nu}\delta h_{\rho \sigma}}\Bigg|_{h=0} 
 \!\!\! 
 +  \frac{\delta \chi_{\delta}}{\delta h_{\mu \nu}} \, C^{\delta \tau } \, 
 \frac{\delta \chi_{\tau}}{\delta h_{\rho \sigma}}\Bigg|_{h=0} 
  . \label{H}
\ee
Following \cite{shapiro3} we can recast (\ref{H}) for the simpler four-dimensional case in the following compact form
\be
&& \hspace{-1.2cm}
\hat{H}^{\mu \nu, \alpha \beta} = \left(  \frac{\omega_{\rm Ric}}{4} g^{\mu (\rho} g^{\nu) \sigma}
- \frac{\omega_{\rm Ric} (\omega_{\rm Ric}+ 4 \omega_{\rm R})}{16 \omega_{\rm R}} g^{\mu \nu} g^{\rho \sigma} 
\right) \times 
 \nonumber 
\Big\{  
\delta^{\alpha\beta}_{\rho\sigma} \Box^{\gamma+2} 
+ V_{\rho\sigma}{}^{\alpha\beta,\lambda_1  \dots \lambda_{2 \gamma +2}} 
\nabla_{\lambda_1}  \cdots \nabla_{\lambda_{2 \gamma+2 }} +
\\
&& \hspace{0.35cm}
+ W_{\rho\sigma}{}^{\alpha\beta,\lambda_1  \dots \lambda_{2 \gamma + 1}} 
\nabla_{\lambda_1} \cdots \nabla_{\lambda_{2 \gamma+1}}  
+ U_{\rho\sigma}{}^{\alpha\beta,\lambda_1 \dots \lambda_{2 \gamma }} 
\nabla_{\lambda_1}  \cdots \nabla_{\lambda_{2 \gamma}} +O(\nabla^{2\gamma-1})  \Big\}     \, , 
\label{acca}
\ee
where 
$\delta_{\mu\nu}^{\rho \sigma} \equiv \delta_\mu^{(\rho}\delta_\nu^{\sigma)} =\frac{1}{2} 
\left(\delta_\mu^{\rho}\delta_\nu^{\sigma}+\delta_\mu^{\sigma}\delta_\nu^{\rho} \right)$, 
and  the tensors $V, W$ and $U$ depend on 
curvature tensors of the background metric and its covariant derivatives.
In (\ref{acca}) the pre-factor in round brackets (called de Witt metric $\cal G^{\mu\nu,\rho\sigma}$)
does not give any contribution to the divergences and, therefore, it
can be omitted. 
The coefficients $\omega_{\rm R}$ and $\omega_{\rm Ric}$ (\ref{acca}) stay for 
$\omega_{{\rm R},\gamma+{\rm N}}$ and $\omega_{{\rm Ric},\gamma+{\rm N}}$  respectively. 
The tensor $V$ is linear in a curvature tensor ($\cal R$), while the 
tensor $U$ contains contributions quadratic in curvature (${\cal R}^2$) and also terms with two covariant derivatives on one curvature ($\nabla^2\cal R$). We  obtain expressions for $U,\,V$ and $W$ tensors by contracting the operator $\hat{H}^{\mu \nu, \alpha \beta}$ with the inverse de Witt metric and extracting at the end covariant derivatives. They have the canonical position of first matrix indices (two down followed by two up) thanks to the application of this metric in the field fluctuation space.

The one-loop effective action is 
defined by 
\cite{shapiro3}
\be
&& \hspace{-0.4cm}
\Gamma^{(1)}[g] = - i  \ln Z[g] = 
\label{Gamma1}
 S_{\rm HD}[g] 
+ \frac{i}{2}   \ln {\rm Det}
(\hat{H}) - i  \ln {\rm Det}(\hat{M})
 - \frac{i}{2} \ln
{\rm Det}(\hat{C})  . 
\ee
Once the relevant contributions to the operator $\hat{H}$ are known we can apply 
the Barvinsky-Vilkovisky method
\cite{GBV} to extract the divergent part of $  \ln
{\rm Det} (\hat{H}^{\mu\nu,\alpha\beta})$. 

The explicit calculation of $\hat{H}$ in a $D$-dimensional spacetime goes beyond the scope of this paper and here we only offer  
the schematic tensorial structure in terms of the curvature tensors of the background metric and its covariant derivatives. 
For the action (\ref{ActionUV2}), where we have only terms with a maximal number of $2\gamma+2\rm N+4$ derivatives on the metric tensor (case of a UV monomial theory), and with front coefficients of these terms given by $\omega_{\rm Ric}$ and $\omega_{\rm R}$,   
the matrix $H_{\mu \nu, \rho\sigma}$ in fully  covariant form  consists solely of 
terms proportional to the 
{\em non-running} constants  $\omega_{\rm Ric}$ and $\omega_{\rm R}$, 
%
\be
&& \hspace{-1.0cm} \hat{H}_{\mu\nu, \rho\sigma} = {\cal G}_{\mu\nu,\alpha\beta}^{-1}\Big( \Box^{\gamma + \mathrm{{N}} + 2} + 
\underbrace{V_{\rho\sigma}^{\alpha\beta( \gamma + \mathrm{{N}} + 2 ) }\,^{\lambda_1\dots \lambda_{2 \gamma + 2 {\rm{N}} +2} }}_{\sim \cal R}\nabla_{\lambda_1} \cdots \nabla_{\lambda_{2 \gamma+ 2 {\rm{N}} +2}} \nonumber \\
&& + 
\underbrace{W_{\rho\sigma}^{\alpha\beta( \gamma + \mathrm{{N}} + 2 ) }\,^{\lambda_1\dots \lambda_{2 \gamma+ 2 {\rm{N}} +1} }}_{\sim \nabla {\cal R}} \nabla_{\lambda_1} \cdots \nabla_{\lambda_{2 \gamma + 2 {N} +1}} 
+ \underbrace{U_{4 \,\rho\sigma}^{\alpha\beta(\gamma+ \mathrm{{N}} + 2 ) }\,^{\lambda_1 \dots \lambda_{2 \gamma + 2 {\rm{N}} } }}_{
\sim {\cal R}^2 + \nabla^2 {\cal R} 
}\nabla_{\lambda_1} \cdots \nabla_{\lambda_{2 \gamma + 2 {\rm{N}}}} 
\nonumber \\
&&
+ \underbrace{U_{5 \, \rho\sigma}^{\alpha\beta( \gamma + \mathrm{{N}} +2 ) }\,^{\lambda_1 \dots \lambda_{2 \gamma + 2 {\rm{N}} -1} }}_{
\sim \nabla {\cal R}^2 + \nabla^3 {\cal R}
}\nabla_{\lambda_1} \cdots \nabla_{\lambda_{2 \gamma + 2 {\rm{N}} -1}} 
+  \underbrace{U_{6 \,\rho\sigma}^{\alpha\beta( \gamma + \mathrm{{N}} + 2 ) }\,^{\lambda_1\dots \lambda_{2 \gamma + 2 {\rm{N}}-2 } }}_{
\sim {\cal R}^3+\nabla^2 {\cal R}^2 +\nabla^4 {\cal R} }\nabla_{\lambda_1} \cdots \nabla_{\lambda_{2 \gamma+ 2 {\rm{N}} -2 }} 
+ \dots 
\nonumber \\
&& 
+ \underbrace{U_{D \, \rho\sigma}^{\alpha\beta(\gamma+ \mathrm{{N}} + 2 ) }\,^{\lambda_1 \dots \lambda_{2 \gamma+ 2 {\rm{N}} +4 - D} }}_{
\sim {\cal R}^{D/2} + \ldots +
\nabla^{D-2} {\cal R}
}\nabla_{\lambda_1} \cdots \nabla_{\lambda_{2 \gamma+2 {\rm{N}} +4 - D}}+O(\nabla^{2 \gamma+2 {\rm{N}} +3 - D})\Big) \,  . 
\ee
To avoid too much complicacy we restricted ourselves above to the case of even dimensionality of spacetime.
We wrote above only terms giving rise to quantum divergences.
We explicitly showed the relationship of the tensors 
$$V^{(i)}, \, W^{(i)}, \, U^{(i)}_4, \, U^{(i)}_5,  \dots, U^{(i)}_D$$
(for the case $i= \gamma+\mathrm{{N}} +2$) 
to the background curvature tensors and its covariant derivatives. 
Employing the universal trace formulae of Barvinsky and Vilkovisky \cite{GBV} 
\be
&& \hspace{2.0cm}
 {\rm Tr} \ln \Box \Big|_{\rm div} \sim \frac{1}{\epsilon} \, \int d^D x \sqrt{|g|} \left( 
  {\cal R}^{\frac{D}{2} } + \nabla^{2}{\cal R}^{\frac{D}{2}-1}+\ldots + \nabla^{D-2} {\cal R} \right),   \\
&& \hspace{-1.2cm}
 \nabla^{p} \, \frac{1}{\Box^{ \mathrm{{N}}  + \gamma + 2}} \, \delta(x,y) \Big|_{\rm div}^{y\to x} 
 \sim   \frac{1}{\epsilon} \, \left(  {\cal R}
^{\frac{p}{2} - (  \gamma+ \mathrm{{N}}  + 2) + \frac{D}{2} } 
+ \ldots + \nabla^{p -   2\gamma- 2\mathrm{{N}}  - 6 + D} {\cal R} \right)
\,\,\,\, \,\, (p \leqslant 2\gamma+2 \mathrm{{N}}  + 4 )   \,  , 
\ee
we can derive in a general case of UV-polynomial theory
the following divergent contribution to the effective action, 
\be 
 &&\hspace{-0.6cm} 
  \Gamma^{(1)}_{\rm div}
 \sim -  \frac{1}{\epsilon}\int d^Dx \sqrt{|g|} 
  \left[ \beta_{\bar{\lambda}} - 2 \beta_{\kappa_D^{-2}} R 
 + \sum_{n=0}^{\mathrm{N} }  \Big( \beta_{\omega_{{\rm R},n}} R  \, \Box^n R 
  +
  \beta_{\omega_{{\rm Ric},n}}   R_{\mu\nu}  \Box^n  R^{\mu\nu}   \Big)  \right. 
  \left.   + \sum_{j=3}^{{\rm N}+2} \sum_{k=3}^{j} \sum_i \beta_{c_{k,i}^{(j)}} \left( \nabla^{2(j-k)} {\cal R}^k \right)_i
 \right] ,  
 \nonumber \\
 &&
\label{Abeta3}
\ee
where all the beta functions 
depend only  on the ``non running"  constants  $\omega_{{\rm Ric},n}$ or $\omega_{{ \rm R},n}$ for $n={\rm N}-\frac{d}{2},\ldots,{\rm N}$. 

\section{Non-local conformal gravity}
\label{NonlocalCG} 
%
%
We here use the compensating field method 
 to make our finite or super-renormalizable gravitational theory ``trivially"
conformally invariant at classical level. 
Following the conventions in \cite{Englert, thooft0}, we replace the following definition, 
\be
g^{\mu\nu} = \left( \phi^2 \kappa_D^2 \right)^{\frac{2}{2 -D}} \hat{g}^{\mu\nu} ,
\label{phighat}
\ee
in the general action (\ref{gravityG}) and we ed up with  
\be
&& \hspace{-0.5cm}
\mathcal{L}_{\rm g} = -  2 \kappa_{D}^{-2} \, \sqrt{g} 
\left( { R(g)} 
+ {\bf Riem(g)}  \, 
{\bf \gamma}(\Box_g) 
{\bf Riem(g)} 
+ {V(g)} 
\right)  
\Big|_{\phi \hat{g}} 
 = -  2  \, \sqrt{ \hat{g} } 
\left[    \phi^2 R(\hat{g}) + \frac{4(D-1)}{ D-2} \hat{g}^{\mu\nu} \partial_\mu \phi \partial_\nu \phi \right] 
\nonumber \\
&& \hspace{0.42cm} 
 -  2 \kappa_{D}^{-2} \, \sqrt{g} \left[ 
{R}({g}) \, 
 \gamma_0( {\Box} )
 {R} (g)
 + {\bf Ric} (g) \, 
\gamma_2( {\Box} )
 {\bf Ric} (g) \right. 
\left. + {\bf Riem}(g)  \, 
\gamma_4({\Box} )
{\bf Riem} (g)
+ {V(g)} \, 
\right] \Big|_{\phi \hat{g}} 
\, . 
\label{Conf1}
\ee
where by $\big|_{\phi \hat{g}}$ we mean that the metric $g^{\mu\nu}$ must be replaced with 
$\left( \phi^2 \kappa_D^2 \right)^{\frac{2}{2 -D}} \hat{g}^{\mu\nu}$. 
The requirement to have a theory completely independent on any scale forses us to identify $\Lambda$ with the Planck mass, namely $\Lambda^2 = (\kappa_D^{-2})^{\frac{2}{D-2}}$. 
The form factors $\gamma_{0,2,4}$ are the same given in (\ref{gamma0}).
We now consider the case $\gamma_4 (\Box) = 0$ in view of having the Schwarzschild spacetime as an exact 
solution of the theory. 
The theory simplify to 
\be
\boxed{\mathcal{L}_{\rm g} 
= -  
2  \, \sqrt{ \hat{g} } 
\left[   \phi^2 R(\hat{g}) +
\frac{4(D-1)}{ D-2} \hat{g}^{\mu\nu} \partial_\mu \phi \partial_\nu \phi \right]  
  - \frac{  2 }{\kappa_{D}^{2}}   \sqrt{g} \left[ 
{ R}({g}) \, 
 \gamma_0( {\Box} )
 { R} (g)
 + {\bf Ric} (g) \, 
\gamma_2( {\Box} )
 {\bf Ric} (g) 
+ { V(g)} \, 
\right] \Big|_{\phi \hat{g}} } \, \,  . 
\label{Conf2}
\ee
In the theory (\ref{Conf2}) we have an extra ghost-like degree of freedom. However, it can be eliminated using the extra symmetry of the theory, namely conformal invariance. 
When the unitary gauge $\phi = {\rm const.}$ is enplaned then we can come back to the 
super-renormalizable or finite quantum theory extensively studied in literatures.  
Moreover, the quantum properties of the theory can not change in different gauges, and the theory must be 
super-renormalizable or finite also when the scalar field is not completely gauged away, but another gauge is implemented. 
We remind that 
for the four dimensional local Weyl theory $\sqrt{|g|} {\bf C^2}$ (\ref{CFM})
the conformal symmetry is usually gauge fixed imposing the graviton fluctuation to be traceless \cite{FradTsi}.

In this theory there are 
``$9+1$"
degrees of freedom (d.o.f.) related respectively to the 
``{$9$}"
metric components of 
$\hat{g}_{\mu\nu}$ and ``{$1$}" scalar field $\phi$\footnote{
For the sake of simplicity we here assume $\kappa_D =1$. 
There is another ``equivalent" way to count the d.o.f. The metric $\hat{g}_{\mu\nu}$ has 
{$10$} d.o.f 
given that it is the spacetime metric, while $\phi$ counts for {$1$} d.o.f for 
a total of {$11$} d.o.f. On the other hand we have one more symmetry (Weyl conformal invariance) that can be used to fix $\phi =1$. Therefore, we end up again with {$10$} d.o.f..
This counting is perfectly consistent with 
the replacement (\ref{phighat}) 
because we have the extra conformal symmetry that we can use to impose for example 
$\sqrt{|\hat{g}}| =1$ or $\phi = 1$
\cite{PercacciPriv, thooft}. 
Indeed, the number of d.o.f. on the left and right side of 
(\ref{phighat}) do match when the gauge freedom is fixed. 
However, in this paper we would like to see (\ref{phighat}) as a splitting of the metric tensor $g_{\mu\nu}$,
having {$10$} d.o.f., 
in {$9+1$} components ({$1$} stays for the conformal overall factor, 
namely $g_{\mu\nu} = e^{2 \omega(x)} \hat{g}_{\mu\nu} \equiv \phi^2 \hat{g}_{\mu\nu}$)
that, afterwards, shows up the new conformal invariance. 
Assuming this point of view, the conformal symmetry can only be spontaneously broken, when we want to avoid losing of one d.o.f. 
%
\\
Let us compare the gravitational Higgs mechanism with an analog toy-model in gauge theory. 
We consider a Lagrangian invariant under local $U(1)$ gauge transformations and based on a real scalar field $\theta(x)$ plus an abelian gauge field $A_{\mu}$ \cite{guada}, 
\be
&& 
{\mathcal{L} = \frac{1}{4} {\bf F(A)}^2 - | D_{\mu } \phi |^2} \, , 
\label{LAphi} \\
&& 
F_{\mu\nu} = \partial_\mu A_{\nu} -  \partial_\nu A_{\mu} \, ,  \\ 
&& D_{\mu } = \partial_\mu - i g A_{\mu} \, ,  \\ \quad 
&&  \phi (x) = \frac{1}{\sqrt{2}} v \, e^{i \frac{ \theta(x)}{v} } \quad (v={\rm const.} \,\, {\rm and} \,\, [v] = 1) \, , 
\ee
where the fields ${\bf A}$ and $\phi$ or $\theta$ transform under $U(1)$ as follows,
\be
&&\phi'(x) =e^{i g \alpha(x) } \phi(x) \quad  {\rm or } \quad \theta'(x) = \theta(x) +  v\,g\, \alpha(x) 
\quad {\rm and} \, , \quad 
  A_{\mu}' = A_\mu + \partial_\mu \alpha(x) \, , \\
&& D_{\mu} \phi =-  \frac{1}{\sqrt{2}} i g \, v \, e^{i \frac{\theta(x)}{v} }
\left(   A_\mu - \frac{1}{v g} \partial_\mu \theta(x)  \right) \, . \label{D2}
\ee
Introducing a new gauge-invariant field 
\be
B_\mu = A_\mu - \frac{1}{v g} \partial_\mu \theta(x) ,
\ee
the Lagrangian turns in 
\be
\mathcal{L} = \frac{1}{4} {\bf F(B)}^2 - \frac{1}{2} g^2 v^2 {\bf B}^2 \, ,
\label{LB}
\ee
which is the Lagrangian for a massive vector field with  {$3$} d.o.f.. Therefore, we start with a theory with  {$3$} d.o.f., {$1$} for the scalar and {$2$} for the massless vector, plus the $U(1)$ symmetry, and we end up 
with a theory containing the same number of d.o.f, but with no more explicit $U(1)$ invariance. 
Similarly, we can gauge fix the $U(1)$ symmetry in (\ref{LAphi}) imposing $\theta(x) = {\rm const.}$, 
and using (\ref{D2}) we get 
\be
| D_{\mu } \phi |^2 =  \frac{1}{2}g^2 v^2 A_\mu A^\mu \, ,
\ee
so that the action (\ref{LAphi}) turns into 
\be
\mathcal{L} = \frac{1}{4} {\bf F(A)}^2 - \frac{1}{2} g^2 v^2 {\bf A}^2 \, , 
\label{LAgf}
\ee
in full agreement with (\ref{LB}) when ${\bf B}$ is identified by ${\bf A}$. One can make the following argument. We have initially {$2$} d.o.f. for the massless vector field and  {$1$} d.o.f. 
for the scalar, but we also have the $U(1)$ symmetry that we can gauge fix to end up 
with only {$2$} d.o.f.. As we have seen this argument is incorrect because the gauge fixing does not eliminates $\theta(x)$, which is actually converted into the longitudinal polarization of the vector field (\ref{LAgf}).  
In gravity the metric plays the role of the massless vector in the above gauge theory example and
the scalar compensator the role of the scalar $\theta(x)$.
In the process of gauge fixing the Weyl symmetry we start with  {$9+1$} d.o.f. 
and we end up with the same number of d.o.f., namely {$10$}, but with $\phi$ absorbed in $\hat{g}_{\mu\nu}$, which 
becomes identical to $g_{\mu\nu}$. In other words in the spontaneous symmetry braking 
process the degrees of freedom can not disappear, but just undergo a redistribution.
\\
Here we did not introduce any potential for the gauge theory as long as for the compensating field 
$\phi$ in gravity. 
Indeed, in both cases the scalar is a spurious d.o.f. that can be completely gauged away.
Moreover, any perturbation $\varphi$ of the scalar $\phi$ around the constant vacuum is confined on the gauge orbit too. Therefore, $\varphi$ is not a physical d.o.f. and can be gauged away. 
One could worry about the stability of the vacuum solution 
$\phi = \kappa_D^{-1}$, $\hat{g}_{\mu\nu} = \eta_{\mu\nu}$.
However, the gravitational theory in the spontaneously broken phase is perturbatively stable as the presence of neither ghosts nor tachyons around the Minkowski vacuum shows 
(see section \ref{propagatorS}.)}. 
However, the product (\ref{phighat}) (and therefore the actions (\ref{Conf1}) or (\ref{Conf2})) are invariant under the rescaling 
\be
\hat{g}_{\mu\nu} \rightarrow \Omega^2(x) \, \hat{g}_{\mu\nu} \, , 
\quad \phi \rightarrow \Omega^{\frac{2-D}{2}}(x) \, \phi ,
\ee
meaning that $\phi$ describes clocks are rulers 
while $\hat{g}_{\mu\nu}$ contains information 
about the light cone causal structure of the spacetime. 
After the gauge fixing $\phi$ is just a constant and the 
distances are now measured by the metric $\hat{g}_{\mu\nu}$ 
or equivalently by the metric $g_{\mu\nu}$. The two metrics are now identified and, therefore,
they have the same number of d.o.f., namely { $10$}. 
In other words the ``effective number of d.o.f." is actually $10$ before 
and after spontaneous symmetry breaking of conformal invariance. The ``Higgs" mechanism 
consist on moving the d.g.f., measuring time and space,  
from $\phi$ to the metric $\hat{g}_{\mu\nu}$. Such degree of freedom is the analog the the Higgs particle
giving mass to the other fields in the standard model of particle physics (SM). 


{\color{black}{\em We infer that the class of finite quantum gravitational theories (\ref{gravityG}) (in odd as well as in even dimension) is actually a range of anomaly-free conformally invariant theories in the spontaneously broken phase of the conformal Weyl symmetry. The conformally invariant theory is given in (\ref{Conf1}) 
or (\ref{Conf2}).}}
%
%
 
 In the gauge $\phi^{\prime} = {\rm constant} = \kappa^{-1}$, one then recover the theory (\ref{gravityG}), 
 and the gauge transformation leading from (\ref{Conf1}) to (\ref{gravityG}) is of course 
$g^{\prime \mu\nu} = g^{\mu\nu} (\phi^2 \kappa_D^2)^{\frac{2}{2-D}}$. 
 
In order to support the claim above at quantum level, in the next section we will prove that the theory is free of conformal anomaly. 

\section{Conformal quantum gravity}
Let us now explicitly prove that the theory is anomaly free, or, which is the same, that the theory is completely finite under quantization. There are two famous examples of conformally invariant gravitational theories in $D=4$, namely
\be
&& \mathcal{L}_1 = a W + \alpha_{\rm v} R^* R \, , \label{CFM}\\
&&  \mathcal{L}_2  = - 12 \phi \left( - \Box + \frac{1}{6} R \right) \phi + 2 \lambda \phi^4 + a W + \alpha_{\rm v} R^* R.
\label{C2FT}
\ee
These Lagrangians are useful tools for understanding the connection between conformal symmetry and the singularities' issue. However, at quantum level they are renormalizable, but not finite, implying that the Weyl symmetry is anomalous. 
In the next subsections we explicitly show that the weakly non-local theory is finite and then free of Weyl anomaly. Therefore, we have a good conformal quantum gravity candidate. 

Besides weak non-locality, another attractive proposal to solve the unitarity problem that plagues the theory (\ref{CFM}) can be read in numerous papers by Mannheim \cite{Mannheim}. 

\subsection{Power-counting renormalizability of (\ref{Conf1})}
We hereby study the divergences of the theory (\ref{Conf1}) when both the scalar 
and tensor perturbations 
propagate, without to take care, for the moment, of the gauge fixing we are going to implement later in the paper. 
To fix the notation, $\hat{h}_{\mu\nu}$ is definite to be the fluctuation respect to 
$\hat{g}_{\mu\nu}$, namely 
\be
\hat{g}_{\mu\nu} = \bar{\hat{g}}_{\mu\nu}+ \hat{h}_{\mu\nu}.
\label{splitghat}
\ee
Let us first to consider the case of the purely scalar interactions. 
We expand the action around a constant dimensionfull background, namely 
\be 
\phi = \kappa_D^{-1} + \varphi \, , \quad {\rm with} \quad [ \varphi] =\frac{D}{2} - 1.
\ee
In the subset of graphs involving only the scalar fluctuation $\varphi$, the kinetic operator and 
the $n-$scalars interactions respectively read 
\be
\mathcal{L}_{\rm Kin, \varphi} \approx \frac{1}{\Lambda^{2 \gamma + D -2}} \varphi \, \Box^{\gamma + \frac{D}{2} } \varphi \, ,  \qquad 
 \mathcal{L}_{\rm int, \varphi} \approx \frac{1}{\Lambda^{2 \gamma + D -2}} 
\frac{\varphi^{n-2} }{\Lambda^{(n-2)\left( \frac{D}{2} - 1\right)}}
\, \varphi \, \partial^{2} \Box^{\gamma + \frac{D}{2} - 1}  \varphi \, , 
\label{scalar}
\ee
with $n \geqslant 3$ arbitrary large.
The power-counting gives,
\be
 \delta^D(K) \, \Lambda^{\left( 1- \frac{D}{2}  \right) E}  \, \varphi^E \, 
\Lambda^{2 \gamma (L-1)} 
 \! \int (d^D p)^L  \left( \frac{1}{p^{2 {\gamma + D}}} \right)^I \!\!  \left( p^{2 \gamma + D } \right)^V 
 =  \delta^D(K) \, \Lambda^{\left( 1 - \frac{D}{2}  \right) E}  \, \varphi^E \, 
\Lambda^{2 \gamma (L-1)} 
 \Lambda_{\rm cut-off}^{D - 2 \gamma (L-1)} \, , 
\ee
where $E$ is the number of external scalar legs, and 
for the scalar sector the relation between $E$, $I$, and $V$ is: $n V = 2 I + E$. 
We end up the with the following superficial degree of divergence for an arbitrary Feynman diagram only involving the scalar field,
\be
\omega_\varphi (G) 
 = D - 2 \gamma (L-1) \, . 
\ee
Notice that the number of external legs is arbitrary because the product $\Lambda^{\left( 1- \frac{D}{2}  \right) E}  \, \varphi^E$ is dimensionless. 
Although, there are only one loop divergences, it seems we have 
an infinite number of operators to renormalize, namely the counterterms read 
\be
\Lambda^{\left( 1 - \frac{D}{2} \right) n } \int d^D x \,  \varphi^{n-2} \varphi \, \Box^{\frac{D}{2}} \varphi \, , 
\ee
where we assumed the form factor to be asymptotically monomial. 
However, we expect these operators exactly to be generated expanding in the fluctuation $\varphi$ 
the curvatures 
$\mathcal{R}^{D/2}$, therefore, the operators we have to subtract 
are only a few when expressed in terms of $\phi$. Such operators can be obtained replacing the metric (\ref{phighat})
in the Lagrangian operators with $D$ derivatives, namely $\mathcal{R}^{D/2}$. 
The operators at the orders $0$, $1$, and $2$ in the curvature are given in the Appendix 
\cite{conformalComp}.  
For ${\bf \hat{R}ic}=0$ from the first two quadratic operators in (\ref{R2}) we get the counterterms we advocated above on the base of the power-counting arguments and only involving the conformal factor.

When all the interactions are taking into account the simpler way to carry out the power-counting 
is using the dimensionless fields $[ \hat{h}_{\mu\nu} ] = [ \varphi ] = 0$. In this case the power-counting is exactly the one given in (\ref{PC}), we only have to rename the graviton fluctuation components 
\be
h_{\mu\nu} \rightarrow \{\hat{h}_{\mu\nu}, \varphi \} \, , 
\ee 
and all the possible counterterms now include also the scalar field $\phi$\footnote{
Notice that expanding the action in $\varphi$ and $\hat{h}_{\mu\nu}$, namely 
$\phi = \kappa^{-1}_D + \varphi$ and 
$\hat{g}_{\mu\nu} = \eta_{\mu\nu} + \hat{h}_{\mu\nu}$,
we get  quadratic mixing terms $(\hat{h}_{\mu\nu} - \varphi)$. Therefore, the quadratic part in 
$\hat{h}_{\mu\nu}$ and $\varphi$ has to be diagonalized by the transformation \cite{Englert}:
\be
\phi \rightarrow \phi + \hat{h}_{\mu\nu} \frac{\delta^2 \mathcal{L}_{\rm g} }{\delta \hat{h}_{\mu\nu} \delta \varphi}\left( \frac{\delta^2 \mathcal{L}_{\rm g} }{\delta \varphi^2} \right)^{-1} .
\ee
}.
Therefore, in $D=4$ and using DIMREG all the possible counterterms of dimension four, up to total derivatives and topological terms, are:
\be
\sqrt{|g|} {R}^2 \, , \quad \sqrt{|g|}{R}_{\mu\nu}^2 \, ,
 \quad \sqrt{|g|}R 
  \, , \quad \sqrt{|g|} = \sqrt{|\hat{g}|}\phi^4, 
\label{counterT}
\ee
in which we have to replace again (\ref{phighat}). 
Notice that the last operator gives rise to the cosmological constant through spontaneous symmetry breaking. However, if we choice a form factor asymptotically monomial only the first two operators are generated by a one-loop computation. 

At this level it is not obvious that the divergences collect all together with the right coefficients to reproduce 
exactly the curvature invariants (\ref{counterT}). Indeed, we will proof that the 
combinatorics are the correct ones later in this section. 
For the case of even dimension, we will make this claim rigorous in the subsection (\ref{C4}) explicitly showing that the quantum dynamics of the fluctuation $\varphi$ is actually trivial because of the
conformal invariance. 

Finally, we want to point out that the main result of this section turns out to be the presence of only one loop divergences. 
Indeed, for the power-counting analysis, given the 
dimensionless fluctuations $\hat{h}_{\mu\nu}$ and $\varphi$, we basically need to look at the number of derivatives in the ultraviolet regime.

\subsection{Conformal quantum gravity in odd dimension}
In this section we show that we fairly easily achieve quantum finiteness in {\em odd dimension} and DIMREG. 
Indeed, in an odd dimensional spacetime there are no one-loop divergences made only of the Riemann tensor because 
we can not construct a curvature invariant out off an odd number of derivatives. 
Other operators involving also the scalar field (like the four operators in (\ref{counterT})) are not generated by the quantum corrections because the form factors are rational functions, namely all the one-loop integrals have the following structure,
\be
 \mathcal{I}_{k,n}  = 
\int d^D p \frac{(p^2)^k}{(p^2+ C )^n} 
= i \frac{C^{\frac{D}{2}-(n-k)}}{(4 \pi)^{\frac{D}{2}}} \, \frac{\Gamma \left( n-k - \frac{D}{2} \right) 
\Gamma \left( k+ \frac{D}{2} \right)}{\Gamma \left( \frac{D}{2} \right) \Gamma(n)}  \,  .
\label{Ikn}
\ee
The above integrals (\ref{Ikn}) are convergent in odd dimension because the gamma function 
\be
\Gamma ( \, \underbrace{  n - k - {D}/{2} }_{ {\rm semi-integer} } \, )
\label{gammaDiv}
\ee
has no poles 
if $n$ and $k$ are both integer, but this is exactly the case of our 
theory, therefore, we do not have one-loop divergences. For $L>1$ the gamma function in (\ref{gammaDiv}) is replaced by
\be
\Gamma \left(  n - k - L \frac{D}{2}  \right) ,
\label{gammaDivl2}
\ee
which can be divergent, for example for $L=2$, because the argument of 
the gamma function is now an integer number.

We end up with a finite and anomaly free theory, which means that the quantum action is conformally 
invariant. 

In the next subsection we will show that the theory in even dimension and in particular in $D=4$ is also quantum finite and anomaly free.

\subsection{Conformal quantum gravity in $D=4$} \label{C4}
In this section we quantize the theory in exactly the same way carried out 
by Fradkin and Tsytlin in \cite{FradTsi} (see also \cite{PercacciConf, Percacci2, Percacci3}) for the local theory (\ref{C2FT})\footnote{
In this subsection we define 
$g_{\mu\nu} \equiv \phi^2 \hat{g}_{\mu\nu}$, which slightly differs from (\ref{phighat})
because $[\phi] = 0$ and in particular its exponent is here independent on the spacetime dimension $D$. 
This definition allows for less cumbersome formulas.}. 

We first introduce the Faddeev-Popov determinant for the conformal symmetry.
The scalar auxiliary field transforms as
\be
\phi^\prime = \Omega^{-1} \phi \equiv e^{ -  \sigma} \phi \approx (1 -  \sigma)\phi 
\quad  \Longrightarrow \quad \delta \phi = -  \sigma \phi \equiv  \xi  R(\phi) 
\quad  \Longrightarrow \quad R(\phi) = \phi \, . \nonumber 
\ee
We split the field in a background $\bar{\phi}$ plus fluctuation $\varphi$ 
and afterwords we impose the gauge condition, namely 
\be
&& \phi = \bar{\phi} + \varphi , \quad \chi (\phi) = \phi - \bar{\phi} = \varphi \, ,  \,\, 
  \mbox{gauge fixing} :
\,\, \varphi = \ell = {\rm const.} .
\ee
Finally, the Fadeev-Popov determinant can be derived as follows, 
\be
&&
\hspace{-0.9cm}
 1 = \Delta_{\rm FP}^{\rm conf}(\phi) \int D \xi \, \delta\left( \chi(^{\xi}\phi) - \ell  \right) 
= \Delta_{\rm FP}^{\rm conf}(\phi) \int D \xi \, \delta\left( \chi(\phi + R \, \xi) - \ell  \right) 
= \Delta_{\rm FP}^{\rm conf}(\phi) \int D \xi \, \delta\left( \chi(\phi) - \ell + \frac{\delta \chi}{\delta \phi} R \, \xi \right) 
\nonumber \\
&&  \hspace{-0.6cm}
=  
\Delta_{\rm FP}^{\rm conf}(\phi) \,  {\rm det}^{-1} \left( \frac{\delta \chi}{\delta \phi} R(\phi)  \right) 
\Big|_{\phi = \bar{\phi}+ \ell} 
 =
\Delta_{\rm FP}^{\rm conf}(\phi) \, {\rm det}^{-1} \left(  \bar{\phi}+ \ell  \right) \,\, 
%
\Longrightarrow \,\, 
\boxed{ \Delta_{\rm FP}^{\rm conf}(\phi) = 
\int D c D \bar{c} \, e^{\frac{i}{2} \bar{c} \ell c + \frac{i}{2} \bar{c} c \bar{\phi}} } \, \,  ,
\label{FPconf}
\ee
where we used the gauge $\chi = \ell$ 
and the determinant is obtained integrating on the anticommutating fields $c,\bar{c}$. 
Since $\ell = {\rm const.}$ the propagator is just a number and all the one loop diagrams are proportional 
to $\int d^4 p \equiv 0$ in dimensional regularization, then $\Delta_{\rm FP}^{\rm conf}[\phi] = 1$\footnote
{
Given the following general operator, 
\be
\mathcal{O} \equiv \mathcal{O}_{0} + \mathcal{O}_{\rm I},
\ee 
where $\mathcal{O}_{0}$ refers to the free part and $\mathcal{O}_{\rm I}$ refers to the 
interaction part, 
\be
&&\hspace{-0.5cm}
 \log {\rm \det{\mathcal{O}}} 
 = {\rm Tr} \log \mathcal{O} 
 = 
{c} + {\rm Tr} [ \log ( 1+ \mathcal{O}_{0}^{-1} \mathcal{O}_{\rm I} ) ]   \label{traccia} \\
&&\hspace{-0.5cm}
 = {c} +  \sum_{n=1}^{+ \infty } \frac{(-1)^{n+1} }{n} \!\!   \int  \!\!
 d^D x_1 \dots d^Dx_n  \,
\mathcal{O}^{-1}_{0}(x_1 - x_2) \mathcal{O}_{\rm I}(x_2)
\mathcal{O}^{-1}_{0}(x_2 - x_3) \mathcal{O}_{\rm I}(x_3)
\dots\dots \mathcal{O}^{-1}_{0}(x_{n} - x_1) \mathcal{O}_{\rm I}(x_1) ,
\nonumber 
\ee
where $c= {\rm constant}$. If the propagator is just a constant and the interactions analytic functions, then 
every integral in the above sum is zero in DIMREG.
}.
Since the scalar fluctuation is just a constant we can redefine the $\bar{\phi} + \ell = \bar{\phi}'$ 
and 
the quantum effective action will only be a function of 
$\bar{g}'_{\mu\nu} = \bar{\phi}^{\prime 2} 
\, \bar{\hat{g}}_{\mu\nu}$ (or $\bar{g}_{\mu\nu} = \bar{\phi}^{2}  \, \bar{\hat{g}}_{\mu\nu}$, which is the same.) 
Let us expand on this point. 

At quantum level we integrate in $\phi$ and $\hat{g}$ with a proper conformally invariant measure
\footnote{
The most famous measures are ultralocal measures that do not depend on derivatives of the metric
tensor $g_{\mu\nu}$. Given two different ultralocal measures they are related by a Jacobian determinant 
that can be formally expressed as the exponential of a $\delta^D(0)$ divergence
\cite{Anselmi:1991wb,Anselmi:1992hv,Anselmi:1993cu}. One possible measure reads
\be
\prod_{\mu \leqslant \nu} \mathcal{D} [ g_{\mu\nu} (- g)^{\frac{k}{2}} ] = 
\prod_{\mu \leqslant \nu} [ \mathcal{D} g_{\mu\nu} ] e^{k \delta^D(0) \int d^D x \log \sqrt{- g}} . 
\ee
The Fujikawa measure is recovered for $k=\frac{(D-4)}{2D}$. However, in DIMREG the value of $k$ 
is immaterial because $\delta^D(0) \equiv 0$ (see for example 4.51 at pag. 43 of reference \cite{GBV})
and we can make a different choice of such exponent 
that will prove useful later in this section.
},
\be
&&
\hspace{-0.4cm} 
Z \! \left( \bar{g}_{\mu\nu} = \bar{\phi}^2 \bar{\hat{g}}_{\mu\nu} \right) =  \! \nonumber \\
&& \hspace{-0.4cm} 
= 
 \int  \! {\mathcal D} \! \left[ \phi (- \hat{g})^{\frac{1}{8}} \right] 
  {\mathcal D} \! \left[ \hat{g}_{\mu\nu} (- \hat{g})^{-\frac{1}{4}} \right] 
 \Delta_{\rm FP}^{\rm diff}(g) \, \delta(\chi(g)^{\alpha} - \ell^\alpha) \, 
  \Delta_{\rm FP}^{\rm conf}(\phi) \, \delta(\chi(\phi)- \ell) \, e^{i S(g)} 
  \Big|_{g_{\mu\nu} = \phi^2 \hat{g}_{\mu\nu}}  
  \label{pathint}\\
  && \hspace{-0.4cm} 
  =  \!\! \int \! e^{\frac{1}{4} \delta^4(0) \int d^4 x \log \sqrt{-\hat{g} }}  \, 
   {\mathcal D} \varphi  \,
  {\mathcal D} \! \left[  \hat{g}_{\mu\nu} (- \hat{g})^{-\frac{1}{4}} \right] 
 {\rm det} (C_{\alpha \beta}(\bar{g}))^{\frac{1}{2} } e^{\frac{1}{2} 
 \chi^{\alpha}(g) C_{\alpha \beta}(\bar{g}) \chi^\beta(g)} 
 \, 
 {\rm det} (M^{\rm diff}_{\alpha \beta}(\bar{g})) \, 
  {\rm det} 
  (\bar{\phi} + \varphi) \, \delta(\varphi - \ell) e^{i S(g)} \Big|_{\phi \hat{g}}  , 
  \nonumber
\ee
where $\Delta_{\rm FP}^{\rm diff}(g)$ and $M^{\rm diff}_{\alpha \beta}(\bar{g})$ have been defined in section \ref{prototype}. 
In  (\ref{pathint}) 
the scalar field $\phi$ and the metric $\hat{g}_{\mu\nu}$ are meant as 
the background fields $\bar{\phi}$ and $\bar{\hat{g}}_{\mu\nu}$ 
plus quantum perturbations, namely 
\be
g_{\mu\nu} = \overbrace{(\bar{\phi}+\varphi)^2}^{\phi^2}\,  ( \, \overbrace{ \bar{\hat{g}}_{\mu\nu} + \hat{h}_{\mu\nu} }^{\hat{g}_{\mu\nu} }  \, ) \, .
\ee 
In (\ref{pathint}) we have multiplied by 
\be
1= {\rm det} (C_{\alpha \beta}(\bar{g}))^{1/2} \int D \ell^\gamma \exp \frac{i}{2} \ell^\alpha C_{\alpha \beta} (\bar{g})\ell^{\beta} \, ,
\ee
and integrated in $\ell^\alpha$. Now we replace in (\ref{pathint}) the conf-determinant (\ref{FPconf})
and we integrate in $\varphi$. However, (\ref{FPconf}) is trivially constant in DIMREG, then 
we can just forget it and make the
choice $\ell = 0$ \cite{FradTsi} to avoid unnecessary constant redefinitions of the background scalar field 
$\bar{\phi}$. Moreover, $\delta^4 (0) \equiv0$ in DIMREG.
The outcome reads
\be
  && Z \left( \bar{g}_{\mu\nu} = \bar{\phi}^2 \bar{\hat{g}}_{\mu\nu} \right) = 
   \int   
  {\mathcal D} \left[ \hat{g}_{\mu\nu} (- \hat{g})^{-\frac{1}{4}} \right] 
 {\rm det} (C_{\alpha \beta}( \bar{g} ))^{\frac{1}{2} } \, 
 {\rm det} (M^{\rm diff}_{\alpha \beta}(\bar{g})) \,
   \, e^{i [ S(g) + S_{\rm gf}(g) 
   ] } \Big|_{g_{\mu\nu} = \bar{\phi}^2 \hat{g}_{\mu\nu}}    \, .
  \label{pathint2}
\ee
Since we completely integrated away the scalar fluctuation, we can now explicitly replace in (\ref{pathint2}) the following simplified splitting, 
\be
g_{\mu\nu} = \bar{g}_{\mu\nu} + h_{\mu\nu} 
= \bar{\phi}^2 (\bar{\hat{g}}_{\mu\nu} + \hat{h}_{\mu\nu} ) = 
{\bar{\phi}}^2 \bar{\hat{g}}_{\mu\nu} + \bar{\phi}^2  \hat{h}_{\mu\nu} = \bar{\phi}^2 \bar{\hat{g}}_{\mu\nu} + 
\bar{\phi}^2  \hat{h}_{\mu\nu} 
= \bar{g}_{\mu\nu} + \bar{\phi}^2  \hat{h}_{\mu\nu}
\, .
\ee 
However, since 
 the measure is conformally invariant it is much simpler to replace the integration in $\hat{g}_{\mu\nu}$ with 
 the one in $g_{\mu\nu}$ and only at the end to replace $\bar{g}_{\mu\nu}$ 
 with $\bar{\phi}^2 \bar{\hat{g}}_{\mu\nu}$. Here are the need steps, 
\be
  &&  
Z \left( \bar{g}_{\mu\nu} 
= \bar{\phi}^2 \bar{\hat{g}}_{\mu\nu} \right) =  \\
&&
= {\rm det} (C_{\alpha \beta}( \bar{g} ))^{\frac{1}{2} } 
 \,
 {\rm det} (M^{\rm diff}_{\alpha \beta}(\bar{g})) \, 
\int   
  {\mathcal D} \left[  \hat{g}_{\mu\nu} (- \hat{g})^{-\frac{1}{4}} \right] 
   \, e^{i [ S(g) + S_{\rm gf}(g) ] } \Big|_{g_{\mu\nu} = \bar{\phi}^2 \hat{g}_{\mu\nu}}    
 \nonumber  \\
 && 
 =  {\rm det} (C_{\alpha \beta}( \bar{g} ))^{\frac{1}{2} } 
 \,
 {\rm det} (M^{\rm diff}_{\alpha \beta}(\bar{g})) 
 \, 
 \left\{ \int   
 {\mathcal D} \left[  {g}_{\mu\nu} (- {g})^{-\frac{1}{4}} \right] 
   \, e^{i [ S(g) + S_{\rm gf}(g) ] }  \Big|_{g_{\mu\nu} = \bar{g}_{\mu\nu} + h_{\mu\nu}}  \right\} \Big|_{\bar{g}_{\mu\nu} = \bar{\phi}^2 \bar{\hat{g}}_{\mu\nu}}
  \nonumber \\
  && 
=  
 {\rm det} (C_{\alpha \beta}( \bar{g} ))^{\frac{1}{2} } 
 \,
 {\rm det} (M^{\rm diff}_{\alpha \beta}(\bar{g})) 
  \left\{  \int e^{-\frac{1}{2} \delta^4(0) \int d^4 x \log \sqrt{- g }}   \,
  {\mathcal D} {h_{\mu\nu}}  
   \, e^{i [ S(g) + S_{\rm gf}(g) ] } \Big|_{g_{\mu\nu} = \bar{g}_{\mu\nu} + h_{\mu\nu}}  \right\} \Big|_{\bar{g}_{\mu\nu} = \bar{\phi}^2 \bar{\hat{g}}_{\mu\nu}}  \nonumber 
   \\
   && 
=   {\rm det} (C_{\alpha \beta}( \bar{g} ))^{\frac{1}{2} } 
 \,
 {\rm det} (M^{\rm diff}_{\alpha \beta}(\bar{g})) 
 \,
\left\{ \int 
  {\mathcal D} {h_{\mu\nu}}  
    \, e^{i [ S(g) + S_{\rm gf}(g) ] } \Big|_{g_{\mu\nu} = \bar{g}_{\mu\nu} + h_{\mu\nu}}  \right\}\Big|_{\bar{g}_{\mu\nu} = \bar{\phi}^2 \bar{\hat{g}}_{\mu\nu}}  
   \label{pathint5}
   \, .
\ee
The divergent contributions to the above one-loop effective action are exactly the ones listed 
in (\ref{counterT}) in agreement with \cite{FradTsi}. The beta functions can be derived as explained in section \ref{prototype} and read out of formula (\ref{Abeta3}) specialized to the case $D=4$.

Therefore, we are quite close to achieve 
conformal invariance at quantum level because the quantum effective action, including all finite contributions, will be a function of $\bar{g}_{\mu\nu} = \bar{\phi}^2
 \, \bar{\hat{g}}_{\mu\nu}$
that keeps correctly hidden conformal invariance. However, we will shortly prove that in order to have  conformal invariance at quantum level the path integral (\ref{pathint5}) must be free of any divergence
in DIMREG\footnote{
If we pretend to keep the Fujikawa measure in the path integral (\ref{pathint5}) the result does not change.
In a $D$-dimesional spacetime the measure reads
${\mathcal D} \left[  \hat{g}_{\mu\nu} (- \hat{g})^{\frac{D-4}{4D}} \right]$, 
but in $D=4$ it is just ${\mathcal D} \, [ \hat{g}_{\mu\nu}] $. Let us now change variable to 
$g_{\mu\nu} = \bar{\phi}^2 \hat{g}_{\mu\nu}$ or actually $\hat{g}_{\mu\nu} = \bar{\phi}^{-2}{g}_{\mu\nu}$,
then the measure turns in 
${\mathcal D} \left[  \hat{g}_{\mu\nu} \right] = {\mathcal D} \left[ {g}_{\mu\nu} \bar{\phi}^{-2}\right]
= \exp \left\{ \delta^4(0) \int d^4 x \log ( \bar{\phi}^{-2}) \right\} {\mathcal D} \left[ {g}_{\mu\nu} \right]
\equiv {\mathcal D} \left[ {g}_{\mu\nu} \right]$ in DIMREG.
%
}.

Now we are in the position to claim 
that the counterterms are the ones given in (\ref{counterT}) and they are conformally invariant. However, since the presence of
\be
\frac{1}{\epsilon} =\log \left( \frac{\Lambda_{\rm cut-off}}{\mu} \right)
\ee
 ($\mu$ is the renormalization scale) in front of each operator in 
(\ref{counterT}), conformal invariance is not preserved at quantum level. Indeed, it is obvious that an ultraviolet cut-off violates local conformal invariance. Equivalently, one may note that the operators 
(\ref{counterT}) are conformally invariant in $D=4$, but not in $4-\epsilon$ because DIMREG does not preserve conformal invariance. Let us name the operators (\ref{counterT}) by $\mathcal{O}_i(g)$,
then when the metric $g_{\mu\nu} = \phi^2 \hat{g}_{\mu\nu}$ is replaced in such operators 
we get the following anomalous contributions to the action, 
\be
\frac{1}{\epsilon} \phi^{- \epsilon} \mathcal{O}_i(\phi^2 \hat{g}; \epsilon) 
\approx \frac{1}{\epsilon} (1 - \epsilon \log \phi ) \, \mathcal{O}_i(\phi^2 \hat{g}; \epsilon) 
=  \frac{1}{\epsilon} \, \mathcal{O}_i(\phi^2 \hat{g}) 
+ \tilde{\mathcal{O}_i}(\phi^2 \hat{g})
-  \log ( \phi ) \, \mathcal{O}_i(\phi^2 \hat{g}) ,
\label{anomaly}
\ee
where the $\epsilon$ among the slots of the operator 
$\mathcal{O}_i(\phi^2 \hat{g}; \epsilon)$ 
comes from the replacement $D \rightarrow D - \epsilon$ in (\ref{R2}). Moreover, the contribution 
 $\phi^{- \epsilon}$ comes from 
 \be
 \sqrt{|g|} = \phi^{D-\epsilon} \sqrt{|\hat{g}|} =  \overbrace{\phi^{D} \sqrt{|\hat{g}|}}^{\rm conf.  \,\, inv.} \, \phi^{-\epsilon} = \sqrt{|g|} \, \phi^{-\epsilon}
 \, . 
 \ee
The last two contributions in (\ref{anomaly}) are finite (independent on $\epsilon$) and explicitly violate conformal invariance. 
The operator $\tilde{\mathcal{O}_i}(\phi^2 \hat{g})$ is the regular contribution to 
$\mathcal{O}_i(\phi^2 \hat{g}; \epsilon)/\epsilon$, namely 
\be
\lim_{\epsilon \rightarrow 0} \frac{\mathcal{O}_i(\phi^2 \hat{g}; \epsilon)}{\epsilon} = 
\tilde{\mathcal{O}_i}(\phi^2 \hat{g})  .
\ee

However, in our theory we have other contributions to the beta functions coming from the potential or 
killer operators. As explicitly shown in \cite{modestoLeslaw} and reminded in section \ref{Theory}
the killer operators contribute to the beta functions linearly in their front coefficients and it is always possible to make the beta functions to vanish. 
Therefore, there is no overall $1/\epsilon$ factor in (\ref{anomaly}) and we can take the limit $\epsilon \rightarrow 0$ consistently with conformal invariance (actually there are no counterterms because the theory is finite.) 
This result is not a fine tuning, but actually one loop exact because the theory is super-renormalizable with no divergences for $L>1$. 

Generalization to any even dimension is 
straightforward once we proved the theory to be finite at quantum level.

\subsection{Evaluating Feynman diagrams}
The most general integral at the $L$-loops order has the following structure,
\be
\int B(k_i, p_i) \prod_{i =1}^L d^D k_i \prod_{j =1}^I \frac{i}{q_j - i \epsilon} \, , 
\label{IntL}
\ee
where $k_i$ are $L$ independent loop momenta, $q_j$ are linear combinations of the $k_i$ and external momenta $p_i$. $B(k_i, p_i)$ is an entire function without poles consisting of the product 
of exponential form factors $\exp - H(q_k)$ coming from the propagators times local and weakly non-local entire functions coming from the vertices, namely
 \be
 B(k_i, p_i) = \prod_{i=1}^I e^{-H(q_j)} \mathcal{I}(k_i, p_i) ,
 \ee
where the entire function $\mathcal{I}(k_i, p_i)$ does not show any pole. 
The general integral (\ref{IntL}) is convergent (up to sub-divergences) for $L>1$ and can be calculated
in Minkowski signature along the real axis because the contribution of the non-local functions to the integrand is even for $k_0 \rightarrow - k_0$ (this is not the case for $\exp  \Box$, which is only convergent in Euclidean signature.) Moreover, the Feynman $i\epsilon$ prescription moves the poles outside the real axis. 
For $L=1$ 
we can integrate in Minkowski signature, as just explained,
whether the integral is convergent. On the other hand any divergent one-loop integral can always be written as the difference of a convergent integral and a divergent rational (by definition it is the ratio of polynomials) integral.
The second one can be evaluated with any technique, with or without making use of the 
Wick rotation, because it is the usual divergence we meet in any local quantum field theory. 

\subsection{Perturbative unitarity}
We can easely address the issue of perturbative unitarity following the original analysis by 
Cutkosky \cite{Cutkosky} and Tomboulis \cite{Tombo, Tomboulis:2015gfa}.
We start by making a non singular field redefinition to bring the kinetic operator of the graviton field 
to be the same of the local Einstein-Hilbert theory, namely 
\be
h_{\mu\nu} \rightarrow \mathcal{F}(\Box) h_{\mu\nu} = e^{- H(\Box)}  h_{\mu\nu}
\ee
 (the field redefinition is not essential whether we do not are interested in formulating the largest time equation.)
Notice that the Jacobian of the transformation is trivially constant (see (\ref{traccia}) in the footnote four)  because the field redefinitions does not involve interactions) 
and all the scattering amplitudes are unchanged since the analytic form factors were all moved in the interaction vertexes \cite{Tombo}. 
Therefore, 
the Landau equations \cite{Eden} for locating the singularities of any given amplitude are not changed by the presence of form factors in the integrand at any loop order. 
Their derivation in \cite{Eden} is the same whether $\mathcal{F}(\Box)$ is a polynomial as in local theories, or a transcendental entire function as in our weakly non-local domain 
(see also formula (\ref{IntL}) in the previous subsection.) Similarly, the derivation of the Cutkosky discontinuity cutting-rules \cite{Cutkosky} is unaffected because it only assumes the   
$\mathcal{F}(\Box)$-factor to be an entire function of 
its argument in any loop amplitude integrand. 
It follows that, at least order by order in the perturbative expansion, the theory is unitary.

\subsection{Unitarity bound and Causality} \label{UnitarityBoundCausality}
In the unbroken phase it is very simple to infer about the unitarity bound and the Bogoliubov-Shirkov causality of the theory.
Indeed, the scattering amplitudes $T_{if}$ (we remind the definition $S=1+ i T$) are zero due to  conformal invariance and in agreement with the Coleman-Mandula theorem, whose hypothesis are here satisfied. In particular it is not necessary that the $S-$matrix is governed by a local theory\footnote{
For a modern proof of the Coleman-Mandula theorem we refer the reader to: S. Weinberg, ``The quantum theory of fields. Vol. 3: Supersymmetry".}. 
Basically, in a conformally invariant theory the scale invariance is added to Poincar\'e invariance, hence physics at different scales is interconnected and the concept of asymptotic states does not make sense anymore 
and the $S$-matrix is trivially the identity. 
Therefore, we may conclude that conformal invariance is a kind of ``strong" achievement of asymptotic freedom. 
Nevertheless, all the non-trivial dynamical nature of the theory is a consequence of 
the spontaneous symmetry
breaking of the Weyl symmetry.  

It deserve to be noticed that an identically zero $T-$matrix in the unbroken phase makes intuitively clear why the theory should be free of classical and quantum spacetime singularities. 
 Indeed, the gravitational collapse is just like a scattering process, but here the 
$S$-matrix is trivial and there is no interaction. Therefore, there is no way to produce a singularity  
in scattering processes, or, which is the same, from the gravitational collapse.
We will expand on this point later from another prospective. 

\section{Spacetime singularities} 
As already pointed out in the abstract and in the introduction, conformal invariance seems to be the unique way 
to get rid out of the spacetime singularities in a gravitational theory. One can quite easily convince himself that in a conformally invariant theory there are no FRW singularities.
Indeed, FRW spacetimes are equivalent, by a conformal transformation, to the Minkowski 
spacetime, which is of course regular everywhere. 
Less trivial is the case of the black hole singularities and numerous attempts have been done in this 
direction \cite{thooft,barsTurok, Bars2}. 

In this section we would complete the studies well displayed and developed 
by Narlikar and Kembhavi \cite{Narlikar} to include also the Schwarzschild metric in their list 
of singularity-free spacetimes. 
Let us first remind the logic introduced by the two authors. 

We start with a Riemannian spacetime manifold $\mathcal{M}$ with a metric tensor $\hat{g}_{\mu\nu}$ and a scalar field $\phi$ as discussed in the previous section. Then we derive the EOM 
for the theory (\ref{Conf2}) that we give here only implicitly for the purpose of this section.
Actually most of the results in this section is true for a general class of
conformally invariant theories, including Einstein conformal gravity given by (\ref{C2FT}) with
$\alpha_{\rm V} =a =0$. 

\noindent
The variation of the action with respect to $\hat{g}_{\mu\nu}$ is:
{\small
\be
&& 
  \frac{ \delta \left[  \sqrt{|g|} \left( R + R 
\gamma_0(\Box) R+R_{\alpha \beta} 
\gamma_2(\Box) R^{\alpha \beta} + { V}
\right) \Big|_{\phi \hat{g}} 
 \right]
}{\sqrt{|{g}_{\phi \hat{g}}|} \, \delta \hat{g}^{\mu\nu}} = 0 \, , 
\ee
\be 
 \hspace{-0.2cm}
  \phi^2 \hat{G}_{\mu\nu}  =
   \nabla_\nu \partial_\mu \phi^2 - \hat{g}_{\mu\nu} \hat{\Box} \phi^2 
 -  4 \frac{D-1}{D-2} \left( \partial_\mu \phi \partial_\nu \phi - \frac{1}{2} \hat{g}_{\mu\nu} g^{\alpha \beta}
 \partial_\alpha \phi \partial_\beta \phi \right) 
 - 
 \frac{ \delta  \left(  \sqrt{|g |}\left( 
  R 
\gamma_0(\Box) R+R_{\alpha \beta} 
\gamma_2(\Box) R^{\alpha \beta} + { V}\right)
\right) \Big|_{\phi \hat{g}} 
}{ \sqrt{| g_{\phi \hat{g}}  |} \, \delta \hat{g}^{\mu\nu} }  .
 \label{EOMg}
  \ee
  }
The variation with respect to $\phi$ is:
{\small
\be
\frac{ \delta \left[  \sqrt{|g|} \left( R + R 
\gamma_0(\Box) R+R_{\alpha \beta} 
\gamma_2(\Box) R^{\alpha \beta} + {V}
\right) \Big|_{\phi \hat{g}} 
 \right]
}{\sqrt{|{g}_{\phi \hat{g}}|}  \, \delta \phi } = 0 
\ee 
\be
  \hat{\Box} \phi = \frac{D-2}{4(D-1)} \hat{R} \phi 
 - \frac{ 1 }{\sqrt{| g_{\phi \hat{g}}  |} } 
 \frac{ \delta  \left(  \sqrt{|g |} 
 \left(
  R 
\gamma_0(\Box) R+R_{\alpha \beta} 
\gamma_2(\Box) R^{\alpha \beta} + {V}
\right)\right) \Big|_{\phi \hat{g}} 
}{ \delta \phi } \, .
\label{EOMp}
\ee
}
\hspace{-0.3cm}
Since all the operators resulting from the variation are at least linear in the Ricci tensor $R_{\mu\nu}$
(notice that this is ${\bf Ric}$ and not ${\bf \hat{R}ic}$), then the Schwarzschild metric is 
an exact solution of the conformally invariant theory, namely 
\be
\!\!\!
g_{\mu\nu} = g_{\mu\nu}^{{\rm Sch}} = 
(\phi \, \kappa_D)^{\frac{4}{D-2}} \, \hat{g}_{\mu\nu} \quad {\rm and} 
\quad \phi = \kappa_D^{-1}  
\,\,\, \Longrightarrow \,\,\, 
R_{\mu\nu}(
(\phi \, \kappa_D)^{\frac{4}{D-2}} \, \hat{g}_{\mu\nu}) = 0 \,\,\, \Longrightarrow \,\,\, E_{\mu\nu}((\phi \, \kappa_D)^{\frac{4}{D-2}} \, \hat{g}_{\mu\nu}) = 0 \, , 
\ee
where by $E_{\mu\nu}$ we mean the set of equations (\ref{EOMg}) and (\ref{EOMp}). 
The EOM are conformally invariant, 
hence if we consider another manifold $\mathcal{M}^*$ obtained from $\mathcal{M}$ by a conformal transformation
\be
 \hat{g}_{\mu\nu}^{*} = \Omega^2 \, \hat{g}_{\mu\nu} \,  ,\quad 
 \phi^{*} = \Omega^{\frac{2-D}{2}} \, \phi \, ,
 \ee
then also $\hat{g}_{\mu\nu}^{*}$ and $\phi^{*}$ satisfy the EOM.
The transformation $\phi \rightarrow \phi^*$ is mathematically valid provided $\Omega^{-1}$ does not vanish (or become infinite). It is assumed that $\Omega=\Omega(x)$ is a twice differentiable function of the spacetime 
coordinates with the demand that always
\be
0 < \Omega < +\infty.
\ee
It is then shown for the Belinskii, Khalatnikov \& Lifshitz (BKL) and for the Taub-Nut metrics that the manifold $\mathcal{M}^*$ is geodesically complete while the original manifold $\mathcal{M}$ is not. Notice that we here changed notation with
respect to the original paper \cite{Narlikar}, namely for us the regular manifold is $\mathcal{M}^*$.

The Schwarzschild metric is an exact solution of the theory (\ref{Conf2}), and can be explictely written in terms of $\phi$ and $\hat{g}_{\mu\nu}$, i.e. 
\be
 g_{\mu\nu}^{{\rm Sch}} = (\phi \, \kappa_D)^{\frac{4}{D-2}} \, \hat{g}_{\mu\nu} \, .
 \ee
 However, as it is evident in the theory we can rescale both the scalar field $\phi$ and the metric $\hat{g}_{\mu\nu}$ to get an
 infinite class of solutions conformally equivalent to the Schwarzschild spacetime, i.e.
 \be
 g_{\mu\nu}^{{\rm Sch}} = (\phi \, \kappa_D)^{\frac{4}{D-2}} \, \hat{g}_{\mu\nu} = 
 (\phi^* \, \kappa_D)^{\frac{4}{D-2}} \, \hat{g}^*_{\mu\nu}
 \quad \Longleftrightarrow \quad 
 \hat{g}_{\mu\nu}^{*} = \Omega^2 \, \hat{g}_{\mu\nu} \,  ,\quad 
 \phi^{*} = \Omega^{\frac{2-D}{2}} \, \phi . 
 \ee
 For $\phi= \kappa_D^{-1}$ and $\Omega=1$ we get the Schwarzschild spacetime 
 $\hat{g}_{\mu\nu}^* = \hat{g}_{\mu\nu}=g_{\mu\nu}^{{\rm Sch}}$. By making use of the conformal rescaling 
 we can construct infinitely many other exact solutions conformally equivalent to the Schwarzschild metric. 
 Moreover, 
 \be
 R_{\mu\nu}(g_{\mu\nu}^{{\rm Sch}}) = 0 \quad \Longrightarrow \quad 
 \hat{R}_{\mu\nu}(\hat{g}^*_{\mu\nu}) \neq 0 .
 \ee
 
We now explicitly provide an example of singularity-free exact black hole solution (in any conformally invariant gravity) obtained by rescaling the Schwarzschild metric by a suitable, singular warp factor $\Omega$ (see also \cite{Prester:2013fia} for use of similar methods). 
 %
For the sake of simplicity here we stay in $D=4$.
Indeed, we regard very educational to include here one regular black hole metric, namely one representative of the gauge conformal orbit, and to study its properties and the spacetime structure. 
The new singularity-free black hole metric looks like (later in this section we will prove the regularity of the spacetime)
\be
&& ds^{* 2} \equiv 
\hat{g}_{\mu\nu}^* dx^\mu dx^\nu = S(r) \hat{g}_{\mu\nu} dx^\mu dx^\nu
= S(r) \left[ \left( 1- \frac{2 m}{r} \right) dt^2 + \frac{dr^2}{1- \frac{2 m}{r}} + r^2 d \Omega^{2} \right] \,  , \label{NRBH} \\
&& \phi^* = S(r)^{-1/2} \kappa_4^{-1} \, . 
\ee
Let us consider the following scale factor $\Omega$ depending only  on the radial Schwarzschild coordinate, 
\be
\Omega^2(r) \equiv S (r) =  \frac{L^4}{r^4}  e^{\frac{1}{2} \left[\log \left( \frac{r}{L}\right)^8 +\Gamma \left(0, \left(\frac{r}{L}\right)^8 \right)  \right]} \, ,
\label{grazieTombo}
\ee
where $L$ is a length scale introduced for dimensional reason, it could be $L = L_P$ (Planck length) or $L=1/\Lambda$ or even $L=2m$. The first two are the scales already present in the theory, while the last one 
is the scale that breaks conformal symmetry on-shell. However, we believe that the spontaneous symmetry breaking of conformal symmetry should happen at the Planck mass scale, therefore, we are led to identify the scale $L$
with $L_P$. 
On the other hand, if we insist on obtaining exactly the Minkowski spacetime for $m=0$, though we are not
forced to do this identification in a conformally invariant theory, and then we can set $L=2m$.

The scale factor $S(r)$ given in (\ref{grazieTombo}) meets $S^{-1}(0) =0$ while $S^{-1}(\infty) = 1$,  
and  
the Schwarzschild singularity appears exactly  where the conformal transformation becomes singular, i.e. where $S^{-1} = 0$. However, the function $S(r)$ is not a gauge-invariant observable in whatever  conformally invariant gravitational theory, therefore, we should not be worried for the singularity present in the transformation law. This is just a gauge artefact and $S(r)$ is not a physical quantity. We have to  worry only about singularities appearing in physical observables. The situation is exactly the same like with the FRW spacetime, where the conformal factor is singular at the time of the Big Bang, but still the conformally equivalent metric is flat and regular everywhere and for any time. 
Notice that the metric $\hat{g}_{\mu\nu}$ that we are tackling in this section has only 
 few non-zero independent degrees of freedom. Indeed, for the Schwarzschild metric we  have only four 
diagonal non-zero components, which are compatible with {$9$}, the maximal number of components of $\hat{g}_{\mu\nu}$.

Of course there is an infinite class of such functions $S(r)$ that enables us to map the singular Schwarzschild spacetime in an everywhere regular one
\footnote{
It is worth to  note here the crucial role played by the entire function (\ref{TomboFF})
once again.}.
 However, as explained at the beginning of this section we must understand the singularity issue just like an artefact of the conformal gauge. 
 
Inter alia, there is a much simpler choice of $S(r)$ respect to (\ref{grazieTombo}) with exactly the same properties,
\be
S(r) = \frac{1}{r^2} \left( \frac{L^4}{r^2}+r^2 \right),
\label{grazieMode}
\ee
and the metric reads 
\be
\boxed{ds^{*2} =  -\frac{1}{r^2} \left( \frac{L^4}{r^2}+r^2 \right) \left( 1 - \frac{2m}{r} \right) dt^2 
+  \frac{1}{r^2} \left( \frac{L^4}{r^2}+r^2 \right) \frac{dr^2}{1 - \frac{2m}{r} } +
\left( \frac{L^4}{r^2}+r^2 \right) d \Omega^{2} } \, .  
\ee
The Kretschmann invariant ${\bf \hat{K} } = {\bf \hat{R}iem}^2$ is also simple and can be displayed here,
{
\be
&& {\bf \hat{K}} = \frac{16 r^2}{\left(L^4+r^4\right)^6} \times \left[ 
 L^{16} \left(39 m^2-20 m r+3 r^2\right)+2 L^{12} r^4
   \left(66 m^2-32 m r+3 r^2\right) \right. \nonumber \\
 && \hspace{3cm} 
  \left. + L^8 r^8 \left(342 m^2-284 m r+63
   r^2\right)+12 L^4 m^2 r^{12}+3 m^2 r^{16}  \right]  .
\ee
}
while the Ricci scalar reads 
\be
\hat{R} = -\frac{12 L^4 r \left(L^4 (r-4 m)+r^4 (3 r-8 m)\right)}{\left(L^4+r^4\right)^3} \, .
\ee
Therefore, ${\bf \hat{K}}$ and $\hat{R}$ are regular for all  $r$.
Finally the Hawking temperature remains unchanged, namely
\be
T_H = \frac{1}{8 \pi  m}.
\ee
We can even consider a more general range of spherically symmetric spacetimes by selecting out 
the following warp factor\footnote{Another scale factor that captures the same properties, but 
will simplify later the analysis of the spacetime geodesic completion, reads as follows, 
\be
S(r) = \left(1 + \frac{L^2}{r^2}\right)^2 .
\label{geoS}
\ee 
Now, the Kretschmann invariant reads, 
\be
{\bf \hat{K} }= \frac{16 \,  r^2 \left[ L^8 \left(39 m^2-20 m r+3 r^2\right)+2 L^6 r^2 \left(42 m^2-16
   m r+r^2\right)+L^4 r^4 \left(150 m^2-108 m r+23 r^2\right)+12 L^2 m^2 r^6+3 m^2
   r^8\right]}{\left(L^2+r^2\right)^8},
\ee
which is regular everywhere and zero in $r=0$. 
},
\be
S(r) = 1+ \left(\frac{L}{r} \right)^\alpha \, \quad \alpha \in \mathbb{N} \, , \quad  \alpha > 2. 
\label{grazieMode2}
\ee
For $\alpha=4$ we get exactly (\ref{grazieMode}), but we explicitly check that the spacetime is singularity-free for any even value of the natural integer $\alpha$.  
For the sake of simplicity we did not consider the metric for general real values of $\alpha$. 
Notice that for the quite general scale factor (\ref{grazieMode2}) 
we get a minimum area of the spatial static sphere centred at the origin 
for $r = 2^{-1/\alpha } \left(\alpha -2 \right)^{1/\alpha } L$. 
The event horizon area and the minimum area are respectively, 
\be
A_H = 4 \pi (2 m)^2 \left[ 1 +  \left(\frac{L}{2m} \right)^\alpha \right] \, , \quad 
A_{\rm min} =  \frac{4 \pi}{4^{ \frac{1}{\alpha }}} 
\left[ 
\left(2^{\frac{1}{\alpha }} \left(\frac{1}{\alpha -2}\right)^{\frac{1}{\alpha
   }}\right)^{\alpha }+1
\right]
\left(\alpha -2 \right)^{ \frac{2}{\alpha} } L^2 \, .
\ee
If we wish to get back the flat spacetime for $m=0$, the natural choice of the scale $L$ turns out to be $L=2m$. Therefore in this case, the event horizon area is $A_H = 32 \pi m^2$ for all $ \alpha$, while minimal area
$A_{\rm min}$ simplifies to 
\be
A_{\rm min} =\pi  \, 2^{4-\frac{2}{\alpha }} \left[ 
  \left(2^{\frac{1}{\alpha }} \left(\frac{1}{\alpha -2}\right)^{\frac{1}{\alpha
   }}\right)^{\alpha }+1
   \right] 
   \left(\alpha -2 \right)^{ \frac{2}{\alpha } } m^2.
\ee
It deserves to be noticed that $A_{\rm min} \leqslant A_H$ for all $ \alpha$ 
and $A_{\rm min} = A_H$ only for $\alpha =4$, which is the minimum value for the parameter $\alpha$ allowed for a singularity-free
spacetime. Moreover, in this case 
a contracting two-dimensional sphere bounces back exactly at the position of the event horizon. 
Finally, for $L=2m$ and $\alpha=4$ the gravitational potential reads
\be
\Phi_{\rm gravity}(r) = -\frac{1+g_{00}}{2}  =- \frac{m}{r} + \frac{8 m^4}{r^4} - \frac{16 m^5}{r^5} .
\ee
At a first sight the choice $L=2m$ seems unacceptable. However, it is exactly the known black hole physics and in particular the trans-Planckian problem that support such identification. Indeed, 
the original derivation of the Hawking radiation involves field modes that, near the black hole horizon, have arbitrarily high frequencies. Therefore, it seems natural that the vacuum for the scalar field $\phi$ is significantly different from  the Planck mass  at such macroscopic quantum scale.  
If we take $L = \beta m$ with $\beta \in [0, +\infty)$, then the area of the event horizon $A_H$ is always bigger that the minimum area $A_{\rm min}$ and only for $\beta =2$ they are equal.

In general relativity we check the spacetimes' regularity by looking at the singularities of Diff.   
invariant operators (the symmetry group is $GL(D$)) constructed out of the curvature and its covariant derivatives. In this way we assess the issue of the presence of curvature singularities only. About the relation of this kind of singularities to the geodesic singularities (which are the subject of powerful Hawking-Penrose theorems in GR) we comment elsewhere. When a spacetime is completely regular, then all invariants must be non-divergent. To prove that a spacetime singularity occurs, it is enough to find one divergent scalar operator. Conversely, in order to prove that no singularity occurs 
(in principle) all invariants should be examined and all should not exhibit curvature singularity in all spacetime points. 
Typically there are infinitely many diff-invariant scalar local operators, that can be constructed, hence the task of proving absence of singularities seems naively impossible. 
However, below we find a nice way out.

In conformal gravity the symmetry group is enlarged to $GL(D$)$\times$Weyl, but still we only need to 
look for singularities in diff-invariant operators. 
Indeed, in $D=4$ there is only one local conformal invariant scalar operator, namely
$\sqrt{|\hat{g}|} {\bf \hat{C}}^2$, 
but it is not diff-invariant. In higher dimensions all conformally invariant scalar invariants are already densitized, so they can not be invariant w.r.t. general coordinate transformation. 
We know that two metrics that differ by a conformal rescaling 
are located on the same conformal gauge orbit. Therefore, there is no physical difference between 
a singular metric and any regular one on the same orbit. 
We only have an operational issue because of the lack of 
scalar invariants under the full symmetry group $GL(D$)$\times$Weyl. 
Therefore, we do not know which invariants to examine, but we can still investigate some operators  
in a subgroup of the full symmetry group $GL(D$)$\times$Weyl. 
Nevertheless, we can easily overcome this problem every time we can explicitly construct the conformal map that 
turns a singular metric into a regular one. 
This is analogous to what was originally done at the event horizon by Kruskal and Szekeres for a Schwarzschild black hole in the Diff-invariant theory.
Once more, the operator $\sqrt{|\hat{g}|} {\bf \hat{C}}^2$ is not a good invariant because it changes under a 
general coordinate transformation (because $\sqrt{|g|}$ is a scalar density). If we insist on using such an operator to investigate the
spacetime structure we are led to claim a persistence of singularity at $r=0$. Indeed, the overall conformal factor 
resulting from the operator ${\bf \hat{C}}^2$ cancels out with exactly the same conformal factor coming from 
the square root of the determinant of the metric. 
For the metric (\ref{NRBH}) 
we get the following chain of identities, 
\be
\sqrt{| \hat{g}^* |}\, {\bf \hat{C}}^2(\hat{g}^*_{\mu\nu}) 
= \sqrt{|\hat{g}|} {\bf \hat{C}}^2(\hat{g}_{\mu\nu})
= \sqrt{|g^{\rm Sch}|} {\bf C}^2(g^{\rm Sch}_{\mu\nu}) 
= \sqrt{|g^{\rm Sch}|} {\bf Riem}^2(g^{\rm Sch}_{\mu\nu}) . \label{chain}
\ee
However, we still have the freedom to make a coordinate transformation to rid of the singularity.
Indeed, the Jacobian resulting from $\sqrt{| \hat{g}^*| }$ plays a similar role to $S(r)$. 

Here we have considered curvature invariants made out of only the metric $\hat{g}_{\mu\nu}$.
However, we can construct an infinite number of operators simultaneously invariant 
under coordinate and conformal transformations when they are built with the metric $\hat{g}_{\mu\nu}$ and 
the scalar field $\phi$. It is sufficient to take any diff-invariant operator of the metric $g_{\mu\nu}$
and to replace in it 
\be
g_{\mu\nu} \quad{\rm by}\quad (\phi^2 \, \kappa_D^2)^{\frac{2}{D-2}} \, \hat{g}_{\mu\nu}.
\ee 
Some examples are given in (\ref{R2}) of the Appendix A. Nevertheless, these operators do not 
 help in understanding the singularities of the spacetime structure, exactly because 
of the presence of the scalar field $\phi$ that can be completely gauged away. 
In other words, in a conformally invariant theory the 
singularity 
is moved from the spacetime metric to the non-physical scalar field. 
Another proposal for an invariant quantity to consider in the case of Diff. and conformal invariant theories is given in the appendix.

We conclude that to understand the spacetime singularities we need to explicitly construct a 
conformal map in such a way, if any, that all diff-invariant operators of $\hat{g}_{\mu\nu}$ are regular everywhere. Therefore, the singularity is non-physical. 

We now apply this procedure to our new metric (\ref{NRBH}), namely we 
evaluate 
the Kretschmann scalar ${\bf \hat{R}iem^2}$ and the Ricci scalar 
for the metric (\ref{NRBH}). 
To avoid cumbersome formulas we only provide the limit of such curvature diff-invariant operators 
near $r=0$,
\be
  \hat{R} = \frac{48 \, m \, e^{ \gamma_E/2 }}{L^4} r + O(r^2) \, ,  \qquad 
  {\bf \hat{R}iem^2} = \frac{624 \, e^{\gamma_E } \, m^2}{L^8} r^2 + O(r^3) \, .
\ee
Plots of the above operators for any value of the radial coordinate are given in Fig. \ref{NLU}.
It deserves to be noticed that for the choice $L \propto m$ the curvature invariants and $T_H$ diverge at the last stage 
of the Hawking evaporation process when $m\rightarrow 0$. 
%

 \begin{figure}
 \begin{center}
  \includegraphics[height=5cm]{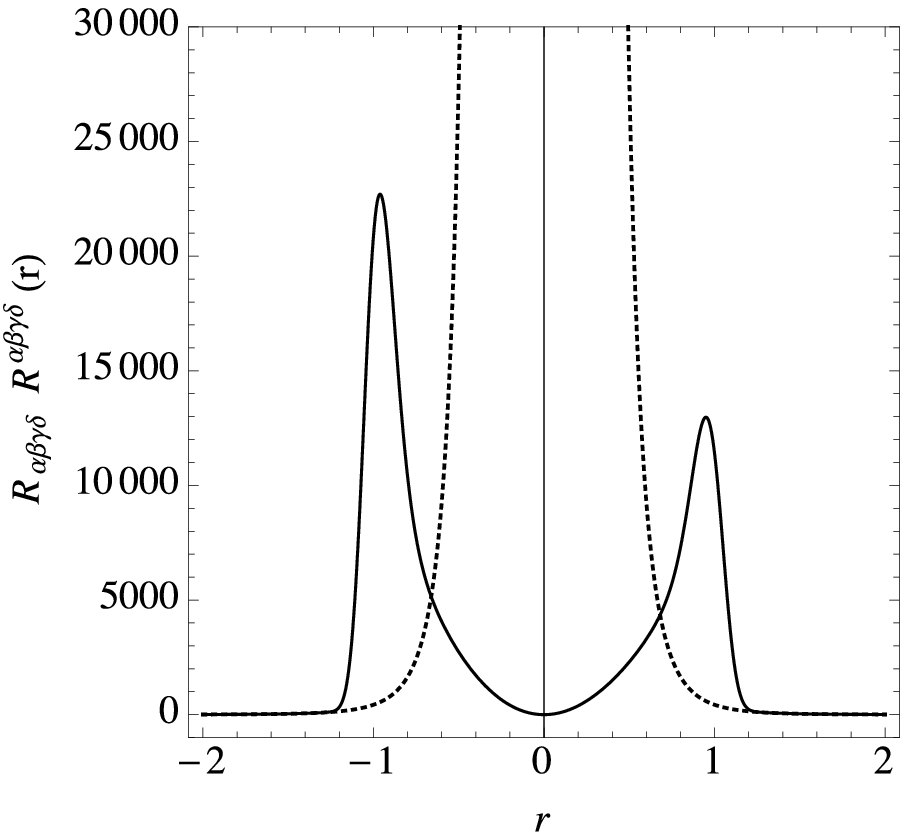}
  \hspace{1.5cm}
  \includegraphics[height=5cm]{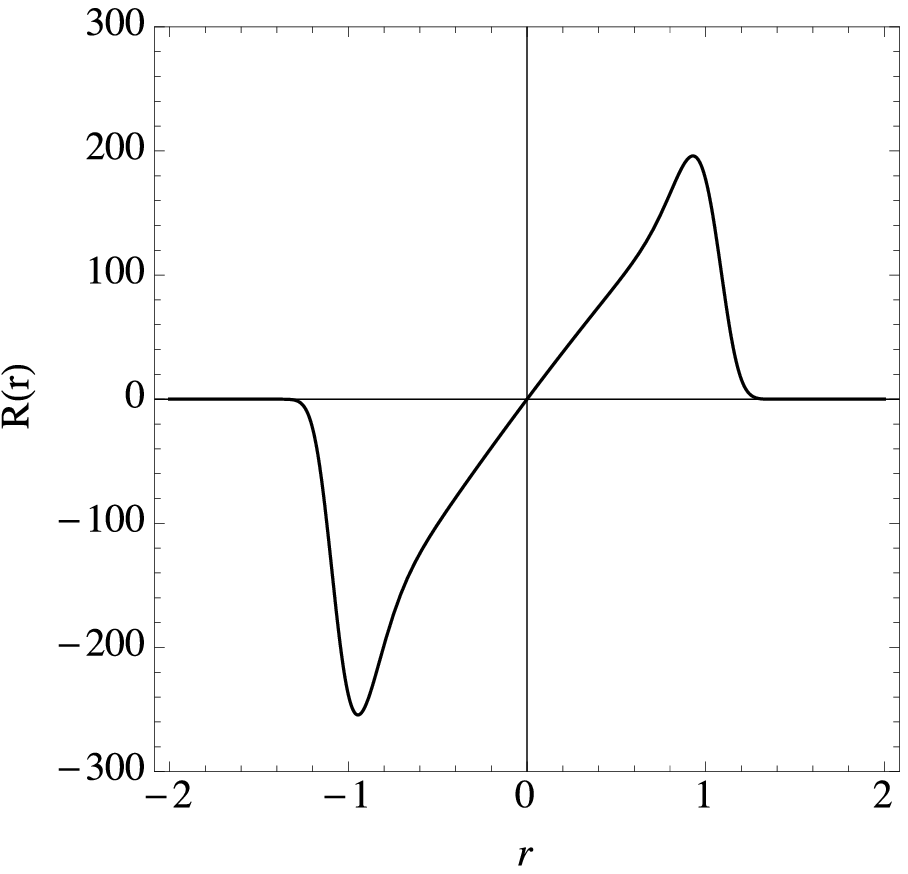}
  \end{center}
  \caption{\label{NLU}The left panel shows the plot of the Kretschmann invariant for $m=3$ and $L=1$. The right panel shows the plot of the Ricci scalar invariant for $m=3$ and $L=1$.   The dashed lines represent the Kretschmann invariant of the original Schwarzschild metric.}
  \end{figure}
%
%
%
%
\begin{figure}

 \begin{center}
 \hspace{-0.5cm}
  \includegraphics[height=5cm]{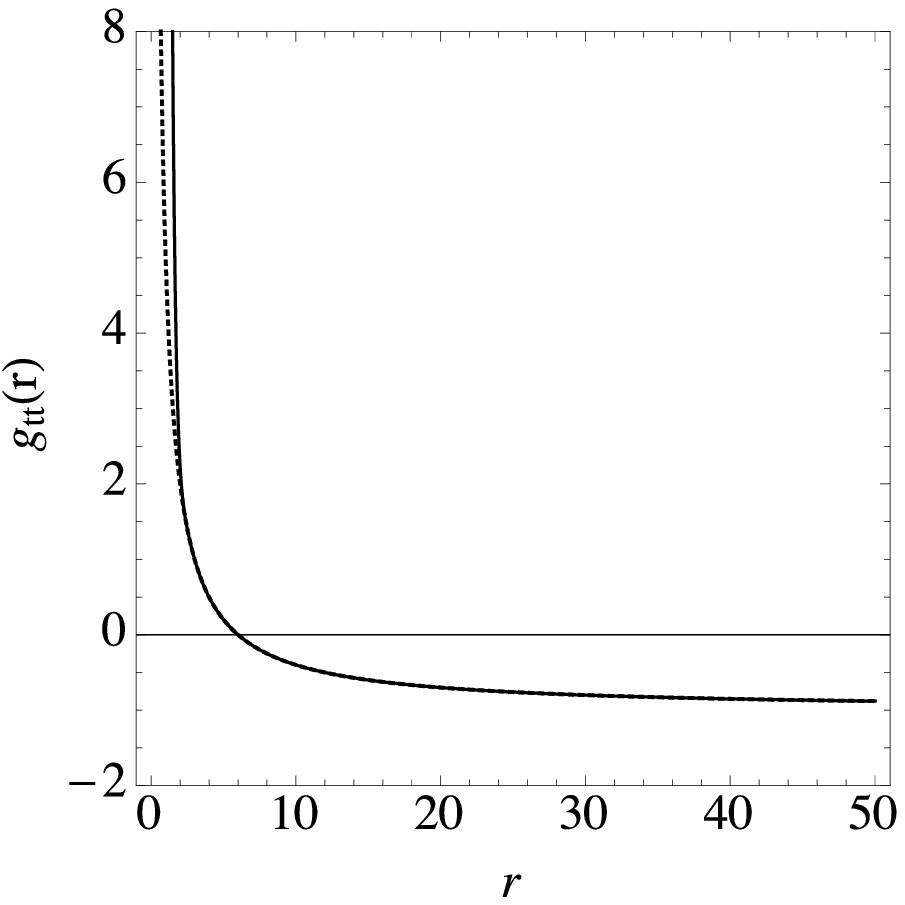}
  \hspace{1cm}
  \includegraphics[height=5cm]{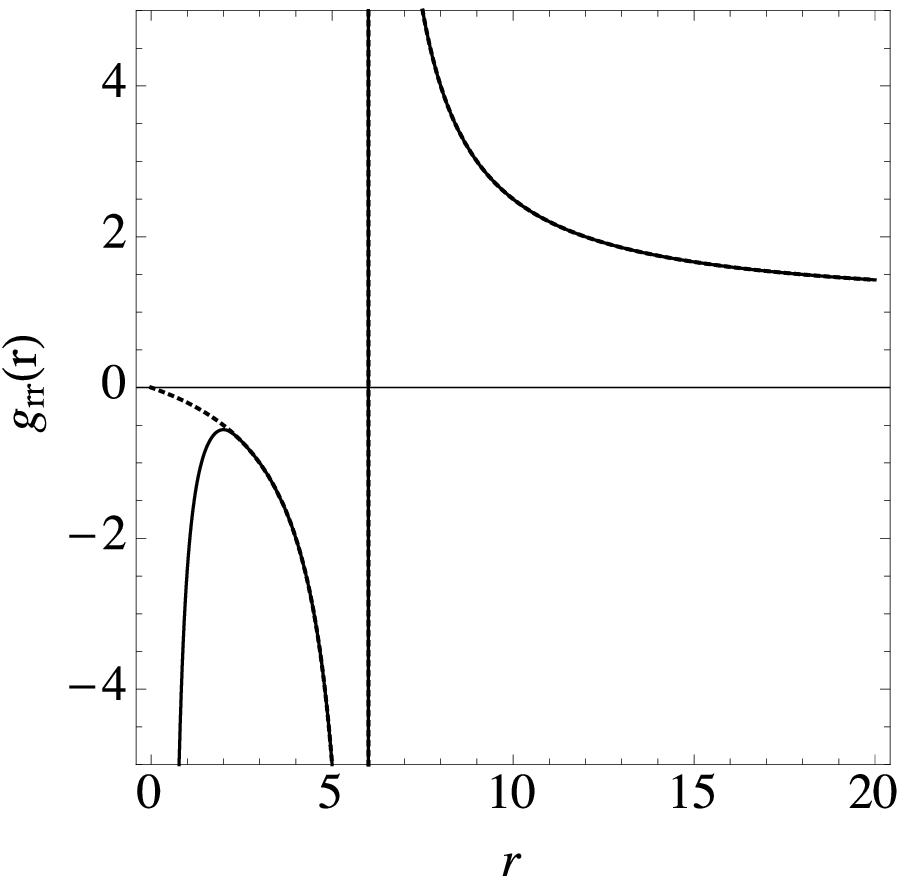}
  \hspace{1cm}
  \includegraphics[height=5cm]{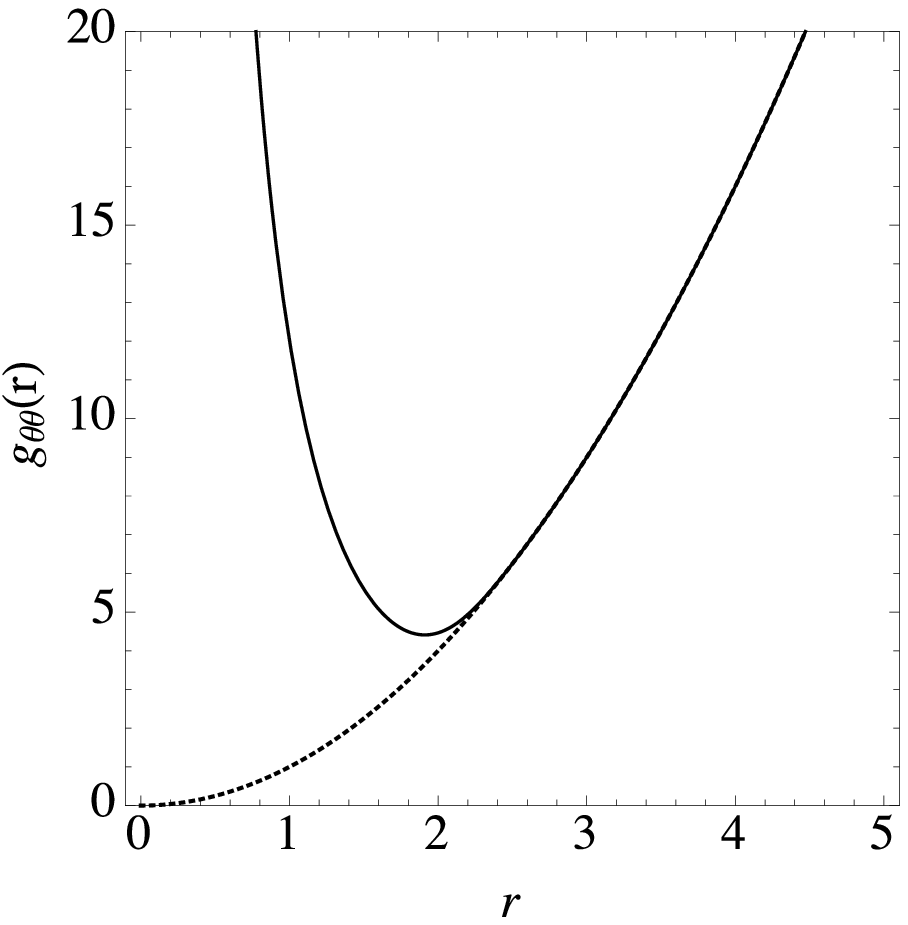}
  \end{center}
  \caption{\label{Metric} 
We have here depicted the components of the metric $g_{tt}$, $g_{rr}$, and $g_{\theta\theta}$.
  The dashed lines represent the components of the Schwarzschild metric. 
  We assumed $m=5$ and $L=1$.  }
  \end{figure}

\begin{figure}
 \begin{center}
  \includegraphics[height=7cm]{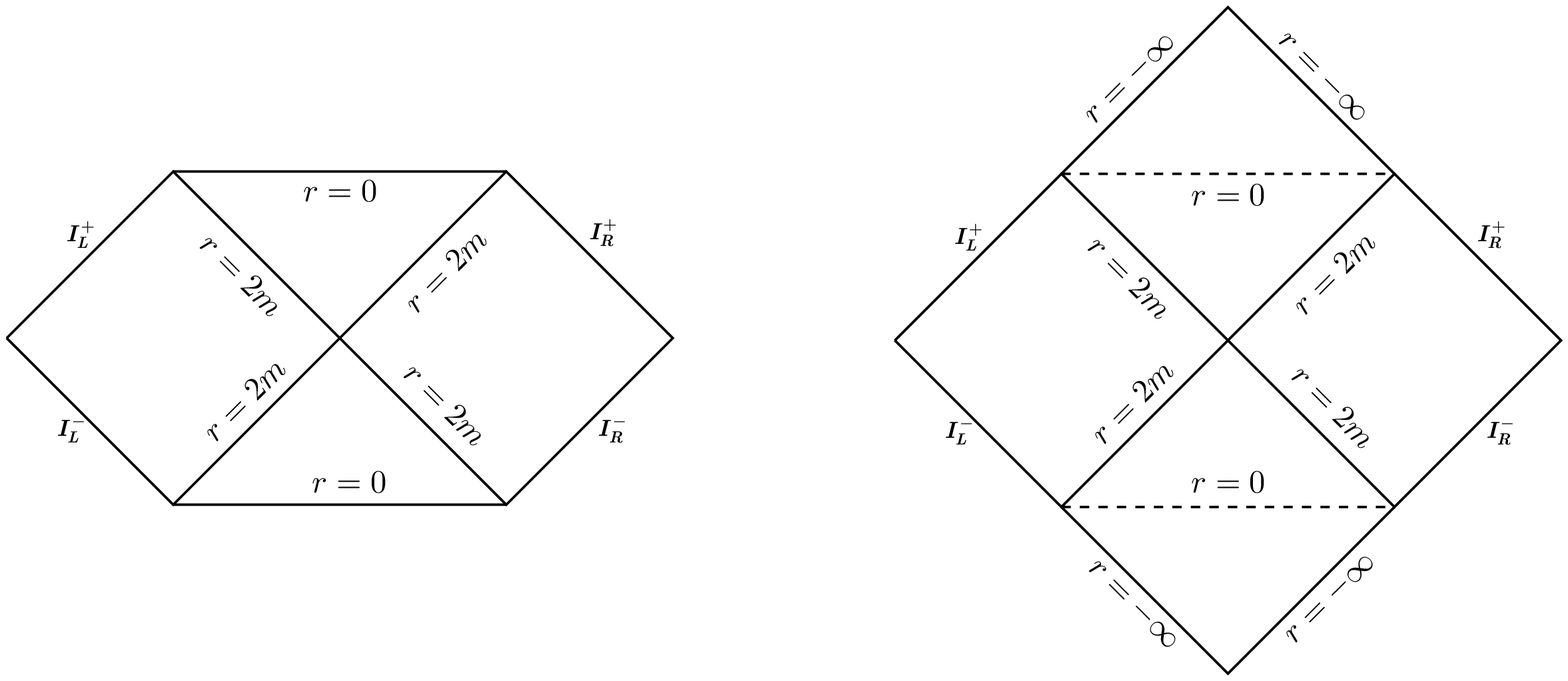}
   \end{center}
  \caption{{Panel on the left.} Spacetime structure of the Schwarzschild metric in conformal gravity.
This diagram has been derived changing coordinates to 
Kruskal-Szekeres ones. The overall  conformal factor $S(r)$ does not change 
the diagram and all the curves $r={\rm const.}$, $t={\rm const.}$, including $r=0$ are located 
in exactly the same positions as in the well known Schwarzschild diagram. 
However, now the spacetime is regular in $r=0$ and the horizontal line can never be reached in finite amount of proper time by any massive particle.
%
If we identify the scale $L$ present in the conformal factor $S(r)$ with $2m$ the 
gauge orbit of the conformal transformations extends to the region $0 \leqslant r \lesssim 2m$.\\
{Panel on the right.} Maximal extension of the singularity-free spacetime accessible only to massless particles. Indeed, the curvature invariants are regular for all $r$ and we can extend the spacetime beyond $r=0$. 
As a matter of comparison we consider the FRW spacetime in Einstein gravity. 
The photon does not see the conformal factor $a(t)$, but we can not extend the
light-like geodesics beyond the Big Bang moment ($t=0$), because $a(0)=0$ and the metric is degenerate in $t=0$. 
Generally, the points beyond which geodesics can not be extended occur as singularities of the curvature invariants. For the new rescaled 
Schwarzschild metric the curvature invariants constructed out of the metric $\hat{g}_{\mu\nu}$
are regular and we are forced to extend the metric to all negative values of the radial coordinate. 
\label{PenroseD}}
  \end{figure}

\section{Geodesic completion A: non-conformally coupled point particle probe}
Having discussed in great extent the curvature singularities, the time has come to touch upon the issue of geodesic completeness of spacetime manifolds in conformal gravity. We will focus on the geodesic motion of some probe material point in the spacetime whose metric is given by \eqref{NRBH}. 
We now show that any probe massive particle can not fall in $r=0$ in a finite proper time.
We will later study the motion of a test point particle conformally coupled to conformal gravity, but the outcome will be essentially the same. 
  We consider the radial geodesic equation for a massive point particle
  \begin{eqnarray}
 (- g_{tt} \, g_{rr}) \dot{r}^2 = E_n^2 + g_{tt},
  \label{geometricabella}  
  \end{eqnarray}
  where  the dot over quantity 
  symbolizes the proper time derivative and $E_n$ is the energy of the point particle. If the particle falls from infinity starting with zero initial radial velocity 
  the energy is the rest energy of massive particle $E_n=1$. 
We can write (\ref{geometricabella}) in a more familiar form 
    \begin{eqnarray}
    \hspace{-0.7cm} 
    \overbrace{(- g_{tt} \, g_{rr})}^{\geqslant 0 \,\, \forall r} 
  \dot{r}^2 + \overbrace{V_{\rm eff}}^{-g_{tt}}(r) = \overbrace{E}^{E_n^2} 
  \quad \Longrightarrow \quad 
   S(r)^2 \dot{r}^2 +S(r) \left( 1 - \frac{2 m}{r} \right) = E \, , \quad  \dot{r}^2 +S(r)^{-1} \left( 1 - \frac{2 m}{r} \right) = S(r)^{-2}E.
  \label{geometricabella2}  
  \end{eqnarray}
Very close to $r=0$ the above equation simplifies to 
  \be
  \dot{r}^2 
  \approx \frac{2 m \, {\rm c}}{L^4} \, r^3 \quad \Longrightarrow \quad
   \dot{r} 
  \approx - \frac{\sqrt{2 m \, {\rm c}} }{L^2} \, r^{3/2} \,  ,
  \label{EG0}
  \ee
  where the numerical constant is ${\rm c}= \exp (- \gamma_E/2)$ for the the metric rescaled by a conformal factor $S(r)$ given by (\ref{grazieTombo}) and
${\rm c}= 1$ for the $S(r)$ in (\ref{grazieMode}). Above we assumed that the particle is falling on the black hole, hence the radial coordinate is decreasing with time $\dot r\leq0$ and this is the reason why the minus sign was chosen. 
  
The plot of $V_{\rm eff}$ can be read out of the plot for $g_{tt}$ in Fig. \ref{Metric}.
From $V_{\rm eff} = - g_{tt}$ we infer that 
any massive particle can arrive in $r=0$.
However, integrating eq. (\ref{EG0}) the proper time to reach the origin $r\to0^+$ turns out to be 
infinite, 
\be
\Delta \tau \approx \frac{2 L^2}{\sqrt{2 m {\rm c} }} \left( \frac{1}{ \sqrt{r} } - \frac{1}{ \sqrt{r_0} } \right)  
\quad \Longrightarrow \quad 
\Delta \tau \equiv \tau(0^+) - \tau(r_0) \rightarrow +  \infty \, . 
\ee
 The maximal extension of the black hole spacetime is given in Fig. \ref{PenroseD}. In short, the Penrose diagram graphically shows that matter never (for none finite time) reaches the point $r=0$. 
 We remind that in the Schwarzschild background a point particle reaches the singularity in finite proper time (see also next section and Fig. \ref{TimeToSing}). To derive the diagram we can write the metric in Kruskal-Szekeres coordinates, namely 
 \be
 ds^{* 2} = S(r(X,T)) \left[ \frac{32 m^3}{r(X,T)} e^{- \frac{r(X,T)}{2 m} }( - dT^2 + dX^2 ) + r(X,T)^2 d \Omega^{2} \right] , 
 \label{PenH}
 \ee
 where $r$ is implicitly defined in terms of $X$ and $T$ through the following equation, $T^2 - X^2 = \left( 1 -r/2 m \right) \exp ({r/2m})$. 

The infinite amount of time needed to reach $r=0$ is a universal property common to all regular spacetimes obtained by applying a conformal analytic transformation to the Schwarzschild metric. 

Let us now evaluate the volume of the black hole interior, namely the volume inside the event horizon. 
For $r<2m$ the radial and time coordinates exchange their role, namely: $r=T$ and $t=R$.
The metric belongs to the class of Kantowski-Sachs spacetimes, 
\be
d s^{*2} = S(T) \left[-  \frac{dT^2}{\frac{2 m}{T} -1}  + \left( \frac{2m}{T} -1 \right) d R^2 + T^2 d \Omega^{2} 
\right] \, , \quad T< 2m \, , 
\ee
and the interior spatial volume reads,
\be
V^{(3)} = 4 \pi  R_o  \,  S(T)^{3/2} \, T^2 \sqrt{\frac{2 m}{T} -1 } 
\ee
that for the choice of the conformal factor $S(r)$ \eqref{grazieMode} turns in  
\be V^{(3)} =
4 \pi R_o  T^2 \sqrt{ \left(\frac{L^4}{T^4}+1\right)^3 \left(\frac{2 m}{T}-1\right)}
 \, , \quad T< 2m \,  ,
 \label{3VSCH}
\ee
\begin{figure}
 \begin{center}
  \includegraphics[height=6.0cm]{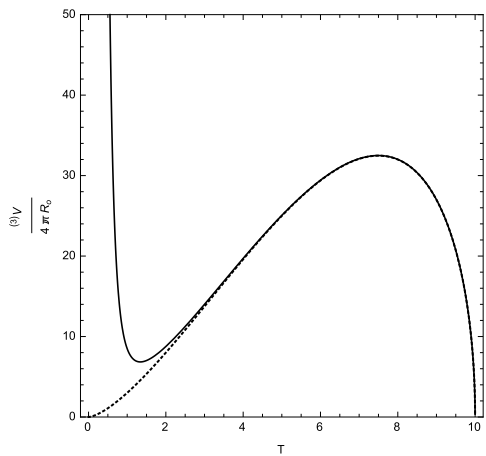}
   \end{center}
  \caption{Plot of the infrared renormalized three-volume $\frac{V^{(3)} }{ 4 \pi  R_o}$ given in (\ref{3VSCH}). We here used $L=1$ and $m=5$.
\label{Volume3}}
  \end{figure}
%
where $ R_o$ is an infrared cut-off due to the translational invariance in the radial variable $R$ of the metric inside the event horizon. 
The volume does not shrink to zero as in the Schwarzschild case, but reaches a minimum value 
and bounces back to infinity for $T \rightarrow 0$ (see Fig.\ref{Volume3}.)

\section{Geodesic completion B: conformally coupled point particle probe}
In this section we study the geodesic completion probing the spacetime with a point particle 
conformally coupled to the Weyl invariant gravitational theory.
The  four dimensional action is obtained replacing again the metric $g_{\mu\nu}$ with $\phi^2\kappa_4^2 \, \hat{g}_{\mu\nu}$
\cite{BekCP}, 
\be
S_{\rm cp} = - \int \sqrt{ - f^2 \phi^2 \hat{g}_{\mu\nu} d x^\mu d x^\nu} 
=  - \int \sqrt{ - f^2 \phi^2 \hat{g}_{\mu\nu} \frac{d x^\mu}{d \lambda} \frac{d x^\nu}{d \lambda} } 
\, d \lambda \, , 
\label{Spc}
\ee
where $f$ is the constant coupling strength, $\lambda$ is a parameter, and $x^\mu(\lambda)$ is the trajectory of the particle. In the unitary gauge $\phi = \kappa_4^{-1}$ the action (\ref{Spc}) 
turns in the usual one for a particle with mass $M = f \kappa_4^{-1}$. 
The Lagrangian reads 
\be
L_{\rm cp} = - \sqrt{ - f^2 \phi^2 \hat{g}_{\mu\nu} \dot{x}^\mu \dot{x}^\nu } \, , 
\ee
and the translation invariance in the coordinate $t$ implies 
\be
\frac{\partial L_{\rm cp}}{\partial \dot{t} } = \frac{f^2 \phi^2 \hat{g}_{tt} \dot{t}}{L_{\rm cp}} = {\rm const.} = E \quad \Longrightarrow \quad \dot{t} = \frac{L_{\rm cp} E }{f^2 \phi^2 \hat{g}_{tt}} . 
\label{ConstE}
\ee
Since we are interesting in evaluating the proper time the particle takes to reach the point $r=0$, we must choice the proper time gauge, namely 
\be
\frac{d \hat{s}^2}{d \lambda^2} = - 1 \quad \Longrightarrow \quad \dot{x}^2 = -1 \, .
\label{PTG}
\ee
Replacing (\ref{ConstE}) in (\ref{PTG}) and using the solution of the EOM for $\phi$, namely 
$\phi = S^{-1/2} \kappa^{-1}_4$ we end up with 
\be
S(r)^2 \dot{r}^2 + S(r) \left( 1 - \frac{2m}{r} \right) - \frac{E^2 \kappa_4}{f^2} S(r) = 0 .
\ee
For a particle at rest at infinity $E = f \kappa^{-1}$ and the above equation simplifies 
to 
\be
S(r) \dot{r}^2 = \frac{2m}{r} \, .
\label{Es}
\ee
For the scale factor (\ref{geoS}) we can easily integrate (\ref{Es}) and 
the evaporation time to reach a general radial position $r$ starting from the event horizon in $r=2 m$ reads
\be
\tau = \frac{4 m^2-3 L^2}{3 m}-\frac{\left(r^2-3 L^2\right) \sqrt{\frac{2 r}{m}}}{3 r}.
\ee
Notice, that for any value of $L\neq 0$ the particle never reach the point  $r=0$,
while for $L=0$ we recover the finite amount of proper time need to reach the singularity in the Schwarzschild metric, namely $\tau_{\rm Sch.} = 4 m/3$ (see Fig.\ref{TimeToSing}.) 
\begin{figure}
 \begin{center}
  \includegraphics[height=7.0cm]{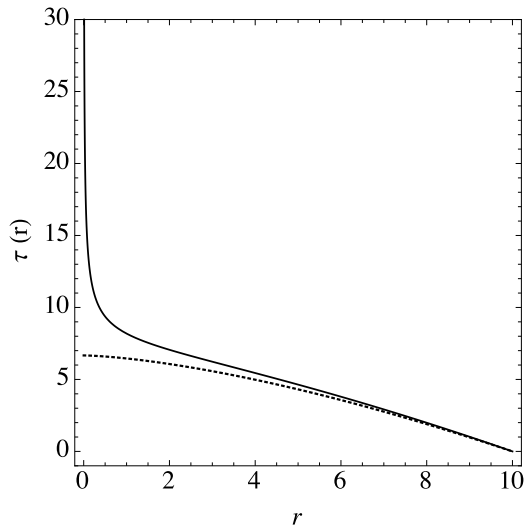}
   \end{center}
  \caption{Plot of the proper time as a function of the radial Schwarzschild coordinate 
  for the regular spacetime and for the Schwarzschild spacetime. We here used $L=2$, to amplify the difference between the two lines, and $m=5$.
\label{TimeToSing}}
  \end{figure}
xdfs
\end{comment}

\section{Belinskii, Khalatnikov, \& Lifshitz singularity}
Another Ricci-flat spacetime is the generalized Kasner universe extensively studied by Belinskii, Khalatnikov \& Lifshitz. It is commonly believed that such spacetime represents the most general 
way a spacetime cosmological singularity can be approached. The line element is 
\be
&& 
ds^{2} = g_{\mu\nu} d x^\mu d x^\nu = -  d t^2 + a_1(t)  d x_1^2 + a_2(t)  d x_2^2+a_3(t)  d x_3^2 \, ,  \nonumber \\ 
&& 
a_i(t) = t^{2 p_i} \quad  (i = 1,2,3) .
\label{BKL}
\ee
The condition to be Ricci-flat ($R_{\mu\nu} = 0$) implies that the metric must satisfy the condition
\be
\sum_i^3 p_i =\sum_i^3 p_i^2 = 1 . 
\ee
The metric (\ref{BKL})
is an exact solution of the theory (\ref{Conf2}) 
as explained just after formula (\ref{EOMp}). Indeed, all Ricci-flat spaces 
(again ${\bf Ric}(g)=0$, but 
${\bf \hat{R}ic}( \hat{g}^*) \neq 0$) are exact solutions of the theory (\ref{Conf2}). 

We now show that in conformal gravity the Kasner singularity is an artefact of the conformal gauge. 
Everything we have to do is to explicitly construct the proper conformal factor that rids of the spacetime singularity. The conformal factor we chose is:
\be
S(t) = \frac{t^2+L^2}{t^2} \, , 
\ee
and the metric reads 
\be
ds^{*2}  \equiv 
\hat{g}_{\mu\nu}^* dx^\mu dx^\nu = 
S(t) ds^2 
= S(t) \hat{g}_{\mu\nu} d x^{\mu} d x^{\nu} = 
\frac{t^2+L^2}{t^2} \left[  -  d t^2 + a_1(t)  d x_1^2 + a_2(t)  d x_2^2+a_3(t)  d x_3^2 \right] \,,
\label{BKLconf}
\ee
where $\hat{g}_{\mu\nu}$ is the BKL metric given in (\ref{BKL}). 
Notice that the metric $S(t) \hat{g}_{\mu\nu}$ is not Ricci-flat, namely 
\be
{\bf \hat{R}ic} (\hat{g}_{\mu\nu}^*=S(t) \hat{g}_{\mu\nu})\neq 0 \, , 
\ee
but ${\bf Ric}({g}_{\mu\nu})=0$ by construction.

For large $t$ the metric \eqref{BKLconf} approaches the metric (\ref{BKL}), but the Kretschmann invariant remains regular,
\be
{\bf \hat{R}iem^2 (\hat{g}^*_{\mu\nu}) } = 
\frac{4 \left(\frac{4 \left(L^2+t^2\right)^4}{u^2+u+1}-\frac{8
   \left(L^2+t^2\right)^4}{\left(u^2+u+1\right)^2}+\frac{4
   \left(L^2+t^2\right)^4}{\left(u^2+u+1\right)^3}+3 L^4 \left(L^4+4 L^2 t^2+8
   t^4\right)\right)}{\left(L^2+t^2\right)^6} \, ,
\ee
where we introduced the usual BKL parametrization, namely 
\be
p_1 =  -\frac{u}{u^2+u+1} \, , \quad p_2 = \frac{ u+1}{u^2+u+1} \quad {\rm and} \quad p_3 =  \frac{u (u+1)}{u^2+u+1}
\, .
\ee
Therefore we proved that after the conformal transformation the BKL metric is without singularity.

\section{Gravitational collapse and Cosmology}

Concerning the black hole singularity problem at quantum level the outcome of Sec.\ref{UnitarityBoundCausality}
may be of some help. Indeed the gravitational collapse (and subsequent Hawking evaporation)
 is just like a scattering process, but the 
$S$-matrix is trivial and there is no interaction in the conformal phase. Therefore, there is no way to produce a singularity. Let us expand on this point  by explicitly constructing the spacetime 
metric for the gravitational collapse in a conformally invariant theory. 
Again, the theory manifests conformal invariance and any metric obtained from the Minkowski one up to a conformal rescaling is an exact vacuum solution. In particular, the metric 
\be
ds^{2} =  \hat{g}_{\mu\nu} d x^{\mu} d x^{\nu} = - d t^2 + 9 \, t^{\frac{4}{3}} d \vec{x}^2,
\label{dust}
\ee
which represents a spacetime filled up with dust in Einstein gravity,
is an exact solution of the weakly non-local theory. It is also a representative metric in the conformal gauge orbit of all conformally flat spacetimes.
Indeed, we can first make the coordinate transformation $t = \tau^3$ to get the metric
\be
ds^{2} = - 9 \, \tau^4 d \tau^2 + 9 \, \tau^4  d \vec{x}^2
\label{dust1}
\ee
and second a conformal rescaling with the factor $S(\tau) = 1/9 \tau^4$ to finally end up with the 
Minkowski flat metric, 
namely 
\be
&& ds^{* 2} = \hat{g}_{\mu\nu}^* dx^\mu dx^\nu = S(\tau) ds^2 = S(\tau) \hat{g}_{\mu\nu} d x^{\mu} d x^{\nu} 
\quad \Longrightarrow \quad ds^{* 2} = 
\frac{1}{9 \tau^4} \left(  - 9\, \tau^4 d \tau^2 + 9 \, \tau^4  d \vec{x}^2  \right) = -  d \tau^2 +   d \vec{x}^2 \, ,  \nonumber \\
&& \phi^* = S(\tau)^{-1/2} \phi = 3 \tau^2 \phi \, .
\ee
Here $\phi^*$ must be constant for consistency when the flat metric 
$ \phi^{* 2} \hat{g}_{\mu \nu}^* =g_{\mu\nu} = \eta_{\mu\nu}$ is an exact solution,
 then $\phi = 1/3 \tau^2$. In other words we do not know what is the
exact solution for $\phi$, but only the conformal relation between 
$\phi$ and $\phi^*$ is known.
Notice that the matter content must be coupled in a conformally invariant way directly in the EOM or in the action, but this is actually irrelevant for the aim of this section. 
 
However, only a  physical source with traceless energy-momentum tensor can be consistently  coupled in a conformally invariant theory
as a consequence of the conformal invariance of the theory. 
Therefore, we can have a dust-like solution not for dust (pressureless) matter, but for collapsing massless radiation. 


We have shown that the Schwarzschild metric is an exact solution in vacuum, while the line element 
(\ref{dust}) 
describes the spacetime inside the black hole filled with matter. 
We can impose the usual boundary conditions to make the metric and the extrinsic curvature continuous 
everywhere, and we finally end up with the simple Oppenheimer-Volkoff model for the gravitational collapse. 
Notice that the solution for $\phi$ is $1/3\tau^2$ in 
the matter region and is independent on time $\tau$ outside, 
but this shall not cause discomfort to the reader because $\phi$ is not  a  physical field. 
However, we advise the reader that we are here dealing with a non-local theory and a
more strict study of the boundary conditions should be required or justified. This last issue is not
present for the theories (\ref{CFM}) and (\ref{C2FT}).

Nevertheless, there is no singularity in $t=0$ because the spacetime (\ref{dust})
is conformally equivalent to the flat one. 
In agreement with the previous section the radiation reaches 
the point $r=0$ and goes beyond, while the regular black hole forms in a finite amount of time.
The Hawking process allows radiation to evaporate out of the black hole after a finite amount of time,
which is invariant independent on the conformal frame 
(because the surface gravity evaluated at the event horizon is invariant under a conformal rescaling of the metric.)
All the radiation is bounces back in our universe in finite time
throughout Hawking particles \cite{thooft} in a way that resembles what we would like to call ``Planck Supernova", 
on the footprint of the proposal 
in \cite{RovelliVidotto} (see also \cite{Barrau, Haggard} and  \cite{DeLorenzo}.)

In a conformally invariant theory  we can introduce a kind of dust matter in a conformally 
invariant way, whenever we have an action principle for dust \cite{Brown, KK}. Indeed, we can make the same replacement (\ref{phighat}) also in the Lagrangian formulation of dust proposed in \cite{Brown} and we trivially achieve conformal invariance of the whole theory. If conformal invariance is spontaneously 
broken, namely $\phi = \kappa^{-1}_D$, then the Lagrangian for ordinary dust matter is recovered. 

Now we are ready to repeat for dust matter the analysis we have just provided for photons. 
The outcome of the previous section is that massive particles reach the point $r=0$ in an 
infinite amount of time, but the regular black hole forms in finite time, just because matter crosses the event horizon in finite proper time. 
Through the Hawking mechanism, particles are emitted to finally make the black hole completely evaporate away.
The dust matter never reaches the origin, but it  bounces back  in a finite amount of time throughout
the process of emission of Hawking particles and they fill back the universe \cite{thooft}. 


We repeated several times that any FRW spacetime is an exact solution of the conformally 
invariant theory whatever is the scale factor $a(t)$. 
The quite involved EOM (\ref{EOMp}) can be solved for the scalar field $\phi$, while the conformal invariance ensures that any solution is pure gauge and then non-physical. We extensively explained 
that the conformal invariance is spontaneously broken and the simpler constant solution for $\phi$ restores the Planck mass (or Newton constant) in the action without any change in the number of d.o.f. of the theory. This is actually a particular vacuum for the scalar field that, however, does not have any dynamics. In this paper we have found other less trivial vacua for the scalar field (see for example $S(r)$) in the black hole case) and in general we expect a general profile for the scalar field condensate
instead of a constant one. 
In other words the EOM for $\phi$ (\ref{EOMp}) can be read as equations for the vacuum of the theory. 
Since any FRW spacetime is an exact solution of the theory ({\ref{Conf2}) or ({\ref{Conf1}) we can infer about the vacuum profile starting from the ``observed" profile for the scale factor $a(t)$, 
namely $\phi^* = a(t)^{-1}  \times {\rm const.}$ It turns out that the FRW background solutions of the theory are consistent with the 
whole spectrum of observations. However, here it is crucial a rigorous analysis of the linear perturbations to select out the physical vacuum. Moreover, the Penrose aeons' theory \cite{Penrose} 
is naturally embedded in any conformally invariant theory, but here we gain the conformal anomaly freedom. 

Finally, we point out again that the evolution of the universe in a conformally invariant 
theory is actually the spontaneous selection of a particular vacuum out of an infinite number of them. 

\section{Conclusions and Remarks}
In this paper we have explicitly shown that a class of finite weakly non-local gravitational theories
is the spontaneously symmetry broken phase of a range of conformally invariant gravities.  
The proof is straightforward in odd dimension, but it has been carry out also in $D=4$ leaving for 
exercise the smooth generalization to any even dimension. 
Since theories are finite then conformal invariance is anomaly free or preserved at quantum level. 
Nevertheless, the conformal invariance is spontaneously broken in exactly the same way like the 
gauge symmetry is broken in the standard model of particle physics (SM). 
Our compensator scalar field $\phi$, need to implement conformal symmetry, plays exactly the same role of the Higgs field. After the symmetry breaking the $\phi$ degree of freedom is absorbed by the spacetime metric that increases its number of components from nine to ten. This is analog to the 
conservation of the degrees of freedom in the SM where three out of the four scalars move to the gauge boson sectors to generate their longitudinal modes. The only difference lies in the absence of any Higgs particle 
in gravity, as a mere consequence of the small number of gauge invariant degrees of freedom respect to the SM.

In the last part of the paper we have explicitly shown how the conformal invariance tames the space time singularities. In particular all the FRW spacetimes, the BKL metric, and the Taub-Nut spacetime are 
singularity-free as shown by Narlikar and Kembhavi 
in \cite{Narlikar}. On the footprint of such masterpiece 
we explicitly proved that 
the Schwarzschild metric is singularity free in a rather general class of conformally invariant gravitational theories. We explicitly construct a class of black hole metrics regular everywhere (including the point $r=0$) that differ from the Schwarzschild metric only for an overall conformal rescaling, which approaches one for distances much bigger then the Planck length or any other scale the conformal factor depends on. 
The conformal map is well defined everywhere and singular in $r=0$ as should be to cancel the singularity. 
The dust dominated universe is also an exact solution and, assuming such spacetime to be the interior of a collapsing star, the simple 
Oppenheimer-Volkoff model for the gravitation collapse turns out be compatible with the collapse 
with a subsequent bounce
\footnote{
It is remarkable the similarity between the class of solution presented in this paper 
and the black hole spacetime structure derived in loop quantum gravity (LQG) in the minisuperspace 
approximation \cite{ModestoSTS} (see also \cite{ModestoLQBH}.). Does LQG show a conformally invariant 
structure at the Planck scale? Preliminary results using weave states \cite{Rovelli} and the ultra-locality of the Hamiltonian constraint are in favour of this interpretation.
}.
Finally, we explicitly showed that the BKL spacetime is singularity-free in non-local conformal gravity as long as in Einstein conformal gravity. 

The main outcome of this paper is the compatibility of conformal invariance with quantum field theory.
Indeed we explicitly constructed a class of conformally invariant gravitational theories free of conformal anomaly at quantum level. Therefore, we overcame the major impediment in believing to conformal invariance as a symmetry realized in nature, although such symmetry must be spontaneously broken. 
In other words the action is invariant under conformal invariance, but the vacuum is not.
When we expand the action around the action around the vacuum $\phi = \kappa_D^{-1}$ 
the action is no more manifestly conformally invariant. In analogy with the Higgs mechanism in the standard 
model, $\phi (x)= \kappa_D^{-1} + \varphi(x)$ plays the role of the forth component of  the scalar field 
$\Phi_4(x) = v + H(x)$. However, here the symmetry is sufficient to completely remove the perturbation 
$\varphi(x)$ from the physcial spectrum, while the spontaneous symmetry braking pattern in the standard model of particle physics, namely $SU(2) \times U(1)_{\rm Y}$ broken to $U(1)_{\rm e.m}$, fixes to zero three out of the four real scalars leaving one dynamical $H(x)$ degree of freedom. 
Notice that the three scalar degrees of freedom of the SM change into the longitudinal 
polarizations of the massive vectors $W^{\pm}$ and $Z^0$, while here the fluctuation $\varphi$ 
is absorbed in the metric $\hat{g}_{\mu\nu}$ (namely $\hat{g}_{\mu\nu} \equiv {g}_{\mu\nu}$) to make it sensitive to the distances and not only the causal 
structure of the spacetime. 





The last comments are about the quasi-polynomial non-local nature of the action. 

The reader could incorrectly objects about the ambiguous presence of form factors and therefore about the predictability of the theory. 
This objection has no physical foundation 
because in the majority of the experiments people are only able to observe form factors. 
If we have the same attitude in QED or QCD etc. then there is no way 
to prove or disprove such theories as long as we only make a finite number of measurements. 
Therefore, our non-local theory is testable on the same level of any other local theory. 

Nevertheless, one could assert that a non-local theory must be the effective theory of some, we would say, ``mystic" extended underlying ``fundamental object". 
Whether we take seriously this point of view then we must find an underlying theory for any 
quantum effective action, which is severely non-local as repeatedly pointed out. However, we are quite happy with the standard model of particle physics (SM) and we are not forced to embed it in a more fundamental theory. However, let us assume for a moment that 
we do it, and we derive the standard model, for example, as a low energy limit sector of string theory. 
This is not enough to make the reductionist reader happy, because the quantum effective action of the string itself will be non-local, or actually double non-local, 
as evident in the string field theory framework when the quantum corrections are included. 
Indeed, the non locality is a feature of any field theory at quantum level and has nothing to do with the 
extended nature of the fundamental constituents. Whatever it is the extended object we base our theory,  
the quantization procedure will introduce extra non locality. Therefore, non locality is a feature 
of quantum field theory.

The difference in our approach is that we start from a classical weakly non-local theory, while in the SM 
all the fundamental interactions are described by a local Lagrangian at classical level. 
However, the theory (\ref{gravityG}) presented in the first section for the minimal case 
$\gamma_4 =0$ is astonishingly local at classical level. Indeed, it has been proved in \cite{Dona} that with a field redefinition we can identically convert, at tree-level, such theory (\ref{gravityG}) in the Einstein-Hilbert action. 
Notice that both the theories have the same perturbative spectrum\footnote{
Let us clarify the perturbative statement with one example. 
In Stelle's quadratic gravity there are other perturbative degrees of freedom, namely one massive spin-2 ghost and one massive scalar, besides the graviton particle. Therefore, the field redefinition argument only applies to the amplitudes with external graviton states. Indeed, these amplitudes coincide with the ones evaluated in Einstein gravity.
In the case of the non-local theory the matching with Einstein gravity is one to one at perturbative level because the perturbative spectrum is the same. However, at the moment the non perturbative spectrum is unknown and the correspondence can not be pushed further.}. 
It is only at one loop that the non locality 
becomes crucial in making the theory finite. However, it is not surprising that the quantum effective action is non-local. As matter of fact any quantum action is non-local whatever the classical action we start from is: local or non-local. In our case we have a kind of ``hidden" non locality present in the classical action that shows up only at quantum level 
{(one can track down the ``hidden" non locality in the non ultra-local path integral measure resulting from the Jacobian of the field redefinition \cite{Dona}.)} 
We infer that our theory is actually local at classical (perturbative) level and non-local at quantum level 
as well as any other theory. The only difference is in a kind of illusory non-locality at classical 
level, which only becomes real at quantum level { when the path integral measure can not be left out.}
Again, at perturbative level we get the same tree-level scattering amplitudes if evaluated in the E-H theory or in the non-local one, which have also the same perturbative spectrum, therefore, the two theories are classically and perturbatively equivalent. 
At quantum level the form factors play a crucial role for the super-renormalizability or the finiteness and the effective action turns out to be non-local.
However, this is not a novelty 
because every theory is non-local at quantum level. 
We can finally firmly assert that the non-local theory is ``{\em actually local}" because of the reasons just explained. 
 
One more comment reads as follows. 
The following operators seem unavoidable at quantum level, 
\be
{\bf Riem}^2 \, {\bf Ric} \, , \quad {\bf Ric} \hspace{-0.8cm}  \underbrace{\mathcal{F}(\Box)}_{\rm non \,\, analytic \,\, function} \hspace{-0.8cm} {\bf Ric} \, , \quad \dots .
\ee
Indeed, they can not be removed by a field redefinition quadratic in the field EOM and/or perturbatively local. 
Moreover, we need the limit $\Lambda \rightarrow \infty$ in (\ref{gravityG}) and in the full quantum action to be convergent. 
However, if one proves that the statements right above are incorrect 
then we can infer that Einstein gravity is finite at any perturbative order and the form factors just play the role of un-physical regulators. 

\section*{APPENDIX A: Curvature operators in conformal gravity}
In this section we remind the outcome of replacing the metric 
$g_{\mu\nu} \equiv \phi^2 \hat{g}_{\mu\nu}$ in the operators linear and quadratic in the curvature. 
In a spacetime of general dimension $D$ the the operators of order zero, one, and two in the curvature read \cite{conformalComp} 
\be
&& \sqrt{|g|} = \phi^D \sqrt{|\hat{g}| } \, , 
\nonumber \\
&& {R} = \phi^{-2} \left [ \hat{R} - 2(D-1)\frac{\Box{\phi}}{\phi} -
(D-1)(D-4) \hat{g}^{ab} \frac{\phi_{,a}\phi_{,b}}{\phi^2}
\right] \, ,  \nonumber \\
&& R^2 
= \phi^{-4} \left[ \hat{R}^2 + 4 (D-1)^2 \phi^{-2} \left( \Box \phi
\right)^2 + (D-1)^2 (D-4)^2 \phi^{-4} \hat{g}^{ab} \phi_{,a} \phi_{,b}
\hat{g}^{cd} \phi_{,c} \phi_{,d} \right. \nonumber \\ 
&& \hspace{0.5cm} - \left.
4 (D-1) \hat{R} \phi^{-1} \Box \phi - 2 \hat{R}
(D-1)(D-4) \phi^{-2} \hat{g}^{ab} \phi_{,a} \phi_{,b} 
+  4(D-1)^2(D-4)
\phi^{-3} \Box \phi \hat{g}^{ab} \phi_{,a} \phi_{,b} \right] \, , \nonumber \\
&& {R}_{ab} {R}^{ab} = \phi^{-4} \left\{ \hat{R}_{ab} \hat{R}^{ab} - 2
\phi^{-1} \left[(D-2)\hat{R}_{ab} \phi^{;ab} + \hat{R} \Box \phi \right] \right. \nonumber \\
&& \hspace{1.2cm}
+ \left. \phi^{-2} \left[ 4(D-2) \hat{R}_{ab} \phi^{,a} \phi^{,b} -
2(D-3) \hat{R} \phi_{,e} \phi^{,e} + (D-2)^2 \phi_{;ab} \phi^{;ab} + (3D -
4) \left( \Box \phi \right)^2 \right] \right. \nonumber \\
&& \hspace{1.2cm}
- \phi^{-3} \left. \left[ (D-2)^2 \phi_{;ab} \phi^{,a} \phi^{,b} -
(D^2 - 5D + 5) \Box \phi \phi_{,e} \phi^{,e} \right] 
+  \phi^{-4} (D-1) (D^2 - 5D + 8) \left(\phi_{,a} \phi^{,a} \right)^2
\right\}  \, . \label{R2}
\ee
Notice that the definition $g_{\mu\nu} \equiv \phi^2 \hat{g}_{\mu\nu}$ differs slightly from (\ref{phighat})
because here $[\phi] = 0$ and its exponent is independent on the spacetime dimension $D$. 

\section*{APPENDIX B: Local and non-local curvature invariants}

In this section we would like to expand on the inadequacy of using conformal invariant operators 
to verify the regularity of the spacetime. 

As already point out in the main text we can construct an infinite number of diffeomorphisms and at the same time conformal invariant operators, whether we make use of both the metric $\hat{g}_{\mu\nu}$ 
and the scalar field $\phi$. Indeed, it is sufficient to take any invariant built with the metric $g_{\mu\nu}$ 
and make the replacement $g_{\mu\nu}= \phi^2 \hat{g}_{\mu\nu}$ (for the sake of simplicity we here in deal with the $D=4$ case.)  
All these operators are very ambiguous because involve the scalar field that can be gauged away exactly by a conformal rescaling. Moreover, these operator do not provide any information about the 
regularity of the physical metric $\hat{g}_{\mu\nu}$ because it is camouflaged with the scalar field. 
Making use of these operators you can only get information about  $g_{\mu\nu}$, but not about 
 $\hat{g}_{\mu\nu}$ solely. Indeed, all this operators diverge exactly as for the Schwarzschild metric. 
However, there is a subclass of conformal invariant operators involving the Bach tensor that has the special property to be ``strongly" regular, namely they are identically zero. One example is:
\be
\phi^{-8} B_{ab} B^{ab} . 
\label{BachCD}
\ee
Moreover, $B_{ab} \equiv 0$ for any FRW and Ricci flat manifold. We could speculate that only vanishing invariants are good conformal invariant operators. Indeed, all the local Diff. invariant operators made of only the metric $\hat{g}_{\mu\nu}$ goes smoothly to zero near $r=0$, while (\ref{BachCD}) is zero everywhere. 
In other words these operators share the same behaviour near the classical singular point. 


Another more speculative idea to check the regularity of a spacetime is related to (strongly) non-local curvature invariants. They arise integrating some invariant densitized scalars
over  aregion of the spacetime volume. We remark that they are not weakly non-local in the sense of containing infinitely many derivatives (such weakly non-local curvature invariants can be constructed as well.) They are integrals over a finite (or infinite) region, so they do not depend only on one point of the manifold. However, the advantage of using them to check the singularities is that they can be at the same time invariants with respect to the group of diffeomorphisms and the conformal transformations group. They are now full invariants, because the densities (coming always with local conformal invariants) are here integrated. One of the example of such non-local invariant in $D=4$ is
\be
X=\int_{V_4}\!d^4x\sqrt{|{g}|} {\bf {C}}^2\,.
\label{NLInv}
\ee
Notice that the local operator $\sqrt{|{g}|} {\bf {C}}^2$ is not invariant under diffeomorphisms as already pointed out in the main text. Namely, it is not a scalar unlike the Kretschmann invariant that we can also 
integrate on a spacetime region, but without to mix together the Diff. properties of the volume element and the curvature invariant.

Thanks to the chain of equalities given in \eqref{chain} the expression $X$ is a conformally invariant non-local scalar and the structural expression for it is the same for any metric 
$g_{\mu\nu}$, ${\hat g}_{\mu\nu}$ or ${\hat g}_{\mu\nu}^*$.

We need to discuss the issue about the domain of integration $V_4$ for such invariant. When a coordinate or conformal transformation is performed the domain of integration must be transformed accordingly. If we are interested in the problem of singularities, then we should choose a spacetime volume that includes the potential singularity point. 
However, despite natural wishes to integrate over the whole manifold, this may be not a good strategy, because the integral may diverge because of the integration on an infinity region. Typically, it is better to restrict the domain to some finite volume near the special point of the manifold where we expect the singularity. For the particular case of the Schwarzschild singularity, which is localized in the coordinate $r$, but not in coordinate $t$ (points $r=0$ are singular for any time $t$) the convenient 4-volume $V_4$ to select is a cylinder. It may 
consist of a finite time interval of coordinate length $\Delta t$ and a 
3-ball $B_3$ of finite radius centred at the origin of spatial reference system. 
We can choose the size of this ball uniformly along the whole cylinder in such a way that its boundary is at coordinate radius $r_1$. For such choice the manifold $V_4$ has the topology of $S_3\times\cal{I}$, where $\cal{I}$ is the interval on the real axis. Since the potential dangerous point is at the origin, we have to cut out a little ball ending at the coordinate position $r_0$ and integrate over the remaining pipe-like shape bulk manifold $\tilde B_3\times\cal{I}$.  

We will investigate the presence of singularities by analyzing the asymptotic behaviour of the invariant $X$, when $r_0\to0$ and keeping $\Delta t$ and $r_1$ fixed and non-zero. The explicit computation for the Schwarzschild metric shows that the densitized invariant $\sqrt{|{g}|} {\bf {C}}^2$ does not depend on $t$, hence the invariant $X$ will be exactly proportional to the interval $\Delta t$. In what follows we can forget about this factor and study the more important $r_0$-dependence. This computation can be performed for any metric $g_{\mu\nu}$, ${\hat g}_{\mu\nu}$ or ${\hat g}_{\mu\nu}^*$. For example in original $g_{\mu\nu}$ we find that near $r=0$ the scaling is the following:
\be
 {\bf {C}}^2=\frac{48m^2}{r^6}\quad {\rm and}\quad \sqrt{|{g}|}=r^2\sin\theta \, . 
\ee
Hence, the invariant $X$ reads
\be
X=\Delta t\int_{\tilde B_3}\!d^3x\sqrt{|{g}|} {\bf {C}}^2=96\pi\Delta t  m^2 \!\int_{r_0}^{r_1}\!\frac{dr}{r^4} \!\int_0^\pi\!\sin\theta d\theta
=-64\pi\Delta t  m^2 \left[\frac{1}{r^3}\right]_{r_0}^{r_1}=64\pi\Delta t  m^2 \left(\frac{1}{r_0^3}-\frac{1}{r_1^3}\right)\,.\label{nlinv}
\label{NL2}
\ee
One can see that $X$ is still divergent, when the internal coordinate radius $r_0$ of the modified ball 
$\tilde B_3$ is sent to zero. The invariant diverges like $64\pi\Delta t  m^2 {r_0^{-3}}$. 
If we believe in the non-local invariant as a good one to identify the singularities,
then the conclusion would be that Schwarzschild singularity is still there even in conformal gravity. 
However, we emphasize that the usage of strongly non-local invariants like $X$ is not so well motivated for the study of singularities. Moreover, we will show later with a counterexample that this invariant 
is in disagreement with the outcome of the explicit study of geodesic completion. 
However, if we take the principal value of $X$ the outcome of (\ref{NL2}) is identically zero because 
the result of the indefinite integral is odd. This is actually in agreement with the value of (\ref{BachCD}),
which is also zero. 
Moreover, it is not in disagreement with all the other Diff. curvature invariants. 

On the other hand the invariant $X$ has quite clear physical interpretation. It is a dimensionless quantity (because of scale-invariance) equal to the value of the conformal gravity action functional evaluated on the portion of the Schwarzschild solution that surrounds the singularity. 
In the result \eqref{nlinv}, we can easily send the external coordinate radius $r_1$ to infinity to end up with the value of the action for the whole Schwarzschild spacetime. In the case of an FRW spacetime the invariant $X$ is clearly zero for any 4-volume because the Weyl tensor in the integrand vanishes. 
It is not surprising that the action functional is a good invariant w.r.t. the Diff. and conformal group. 
The only problem is that it is non-local by construction. 
Requiring that it is well-defined and vanishes on the flat spacetime in $D=4$ we have a unique choice for it, which is the familiar action of the $4$-dimensional conformal gravity. In higher dimensions we can have more choices for conformally invariant actions satisfying the above condition, however, still finitely many. As it is known only non-local global invariants (like ADM mass for example) are true observables in pure classical GR. The invariant $X$ is one of such invariant for a 4-dimensional conformally invariant gravitational theory. A similar invariant can be constructed in conformally invariant 4-dimensional electrodynamics, namely 
\be
X'=\int_{V_4}\!\sqrt{|{g}|} {\bf {F}}^2\,, 
\ee
where $\bf F$ is the field strength of electromagnetism. We find that for the Coulomb potential 
$\Phi_{\rm el}=Q/r$, the non-local invariant is also divergent $X'=8\pi\Delta t Q^2 r^{-1}$ and it measures the energy stored in the field near the point-like charge 
multiplied by the time interval $\Delta t$. 
In this case, as for conformal gravity, the singularity is present because of the infinite energy stored in the gravitational field near the singularity.

We notice that the non-local invariants are identically zero whether we shrink to zero the integration 
domain, namely 
\be
\lim_{r_0 \rightarrow r_1} X = \lim_{t_0 \rightarrow t_1} X =0.
\ee
A potential solution of this ambiguity is to define the invariant $X$ by taking the principle value of the
integral, namely
\be
{\rm P} 
( X )= {\rm P} \left( \Delta t\int_{\tilde B_3}\!d^3x\sqrt{|{g}|} {\bf {C}}^2 \right) 
=
-64 \pi\Delta t  m^2 \, {\rm P} \left(\frac{1}{r^3} \right) = 0
\,.\label{nlinv2}
\ee
Another drawback of using conformally invariant operators is related to the fact that conformal symmetry is in the broken phase for all the solutions analyzed here.

The arguments presented 
in this section should persuade the reader that 
 we can not infer about the regularity of the spacetime 
in a conformally invariant theory by the meaning of appropriate singularity-free curvature operators that are invariant under both conformal and general coordinate transformations. 
Only the Diff. invariant operators made of $\hat{g}_{\mu\nu}$ are appropriate to 
infer about the singularities of the spacetime, namely
\be
{\bf \hat{R}(\hat{g})} \, , \quad {\bf \hat{R}^2(\hat{g})} \, , \quad
{\bf \hat{R}ic(\hat{g}) \, \hat{R}ic(\hat{g})} \, , \quad {\bf \hat{R}iem(\hat{g}) \, \hat{R}iem(\hat{g})} \, , \, \dots \, . 
\label{diffinv}
\ee
Let us expand on this latter statement. 

In this section we presented two different kinds of operators.
A first range of operators involve not only the metric, but also the scalar field. These operators are not able to single out the spacetime properties of the metric $\hat{g}_{\mu\nu}$. 
They are only sensitive to the metric $g_{\mu\nu}$, but not to the sub-spacetime structure
made of $\phi$ and $\hat{g}_{\mu\nu}$. 
A second range of operators involve only the metric $\hat{g}_{\mu\nu}$ (and not $\phi$), 
but are non-local and not suitable 
to infer about the regularity of the spacetime because the geodesic completion has to do 
with the local properties of the manifold.


Finally, suppose that the regular metric $\hat{g}^*_{\mu\nu}$ for a spherically symmetric body 
is not a solution of any conformally invariant theory. Therefore, nobody can object whether we use 
any local curvature invariants (\ref{diffinv}) to infer about the regularity of the spacetime.
Indeed, it turns out that the metric $\hat{g}^*_{\mu\nu}$ is singularity-free, because all the invariants 
(\ref{diffinv}) are regular everywhere. 
On the other side the spacetime is geodesically complete as explicitly shown in the text.
Therefore, the outcomes of the Diff. invariant operators and of the geodesics completion perfectly agree. 
On the other hand if we use the non-local operator we find a singularity in the curvature invariant,
while the geodesic motion is well defined. 
This argument is sufficient to rule out the non-local operator (\ref{NLInv}) as a good tool to probe the
spacetime singularity structure, because  it is in disagreement with the geodesic completion. 
However, in this case the metric is not solution of any conformally invariant theory and in principle there is no reason to check the regularity of the spacetime by using general conformal invariant operators.


Therefore, we could infer that only the local curvature invariants (\ref{diffinv}) made of the metric 
$\hat{g}_{\mu\nu}$ solely can eventually point out the presence of singularities in the spacetime. 


\end{document}